\documentclass[useAMS,usenatbib]{mn2e}
\usepackage{color,hyperref}

\usepackage{float}
\usepackage{xcolor}
\usepackage{graphicx}
\usepackage{multirow}
\usepackage{amsmath}
\usepackage{amssymb}

\usepackage{txfonts}
\DeclareSymbolFont{cmletters}{OML}{cmm}{m}{it}
\DeclareMathSymbol{v}{\mathalpha}{cmletters}{"76}

\newcommand{\tr}[1]{\textrm{#1}}
\newcommand{\trt}[1]{\textrm{\tiny{#1}}}

\newcommand{\msol}{\tr{M}_{\odot}}
\newcommand{\rsol}{\tr{R}_{\odot}}

\newcommand{\mbh}{M_\bullet}

\newcommand{\mstar}{M_\star}
\newcommand{\rstar}{R_\star}

\newcommand{\E}[1]{\times 10^{#1}}
\newcommand{\lr}[1]{\left(#1\right)}
\newcommand{\scale}[3][]{\lr{ \frac{#2}{#3} }^{#1}}

\newcommand{\sw}{Sw1644+57}
\newcommand{\tchek}{TCH14}

\newcommand{\tsec}[1]{\textsection\ref{#1}}
\newcommand{\movielink}[1]{\href{#1}{Animation on YouTube (#1), and online.}}

\newcommand{\tm}{T_\trt{M}}
\newcommand{\rt}{R_\trt{T}}
\newcommand{\rp}{R_\trt{P}}
\newcommand{\major}{a_\trt{M}}
\newcommand{\minor}{b_\trt{m}}
\newcommand{\mfive}{10^5 \, \msol}
\newcommand{\mbhfive}{ {\mbh}_5 }
\newcommand{\mstarz}{ {\mstar}_0 }
\newcommand{\ttime}{ \tau }
\newcommand{\lumfe}{ L_{48} }
\newcommand{\rs}{ R_{\rm S} }
\newcommand{\rrad}{ r } 
\newcommand{\kkh}{k_\trt{KH}}
\newcommand{\krt}{k_\trt{RT}}
\newcommand{\bstar}{B_\star}
\newcommand{\bstr}{B_\trt{str}}
\newcommand{\strdens}{\rho_\trt{s}}

\newcommand{\pbeta}{\beta_\trt{p}}
\newcommand{\bstarz}{B_{\star 0}}
\newcommand{\rstr}{R_\trt{str}}

\newcommand{\eqsp}{\,\,\,\,}
\newcommand{\bfac}{f_{-3}}
\newcommand{\ximo}{\xi_{-1}}
\newcommand{\xihalf}{\xi_{1/2}}

\title[Magnetic Flux Capture by Tidal Debris Stream]{Tidal Disruption and Magnetic Flux Capture: Powering a Jet from a Quiescent Black Hole}
\author[L.Z.~Kelley et al.]{Luke Zoltan Kelley$^{1}$\thanks{E-mail:
lkelley@cfa.harvard.edu} , Alexander Tchekhovskoy$^{2,3}$ and Ramesh Narayan$^{1}$
\\
$^{1}$ Harvard University, Center for Astrophysics, Cambridge MA 02138 \\
$^{2}$ Department of Physics and Department of Astronomy, University of California, Berkeley, CA 94720-3411\thanks{Einstein Fellow}\\
$^{3}$ Lawrence Berkeley National Laboratory, 1 Cyclotron Rd, Berkeley, CA 94720\\
}

\begin{document}

\date{Accepted: 2014 September, 30. Received: 2014 July, 22}
\pagerange{\pageref{firstpage}--\pageref{lastpage}} \pubyear{2014}

\maketitle
\label{firstpage}

\begin{abstract} 
The transient Swift J1644+57 is believed to have been produced by
an unlucky star wandering too close to a supermassive black hole (BH) leading to a 
tidal disruption event.  This unusual flare displayed highly super-Eddington X-ray emission
which likely originated in a relativistic, collimated jet.
This presents challenges to modern accretion and jet theory
as upper limits of prior BH activity, which we obtain from the radio afterglow of this event,
imply that both the pre-disruption BH and stellar magnetic fluxes
fall many orders of magnitude short of what is required to power the observed X-ray luminosity.
We argue that a pre-existing, ``fossil'' accretion disc can contain a sufficient reservoir of magnetic flux
and that the stellar debris stream is capable of dragging this flux into the BH.
To demonstrate this, we perform local, 3D magnetohydrodynamic simulations of
the disc--stream interaction and demonstrate that the interface between the two is unstable to
mixing.  This mixing entrains a sufficient amount of fossil disc magnetic flux into the infalling
stellar debris to power the jet. We argue that the interaction with the fossil disc can
have a pronounced effect on the structure and dynamics of mass fallback and likely the resulting
transient. Finally, we describe possible ramifications of these interactions on unresolved problems
in tidal disruption dynamics, in particular, the efficiency of debris circularization, and effects
of the disruption on the preexisting black hole system. \\
\href{http://goo.gl/T84tLs}{Animations online: http://goo.gl/T84tLs}
\end{abstract}

\begin{keywords}
MHD, stars:black holes, galaxies: jets, quasars: supermassive black holes, galaxies: kinematics and dynamics
\end{keywords}

% = = = = = = = = = = = = = = = = = = = = = = = = = = = = = = = = = = = = = = = = = = = = = = = = = = 
% = = = = = = = = = = = = = = = = = = = = = BEGIN DOCUMENT = = = = = = = = = = = = = = = = = = = = = =
% = = = = = = = = = = = = = = = = = = = = = = = = = = = = = = = = = = = = = = = = = = = = = = = = = = 

% ================================================================
% ========================= INTRODUCTION =========================
\section{Introduction}

% ==== Introduction: Tidal Disruption =====
\subsection{Tidal Disruption}
The same disparity in gravitational force across a body which causes the ocean tides on earth can become strong enough to destroy a star if the gravity source is sufficiently massive and dense \citep{hill75,fran76,youn77}.  Such a \emph{tidal disruption event} \citep{rees88} is possible when a star passes within the \emph{tidal radius}, which is defined as the point at which tidal forces overcome the self-gravity of an object, or equivalently the radius at which the dynamical time of the orbit matches that of the star:
	\begin{equation}
	\rt \equiv \rstar \scale[1/3]{\mbh}{\mstar},
	\end{equation}
where $\mbh$ is the BH mass and $\mstar$ the stellar mass.
The tidal force decreases as $\mbh/r^{3}$, but the Schwarzschild radius is related to mass as $\rs = 2G\mbh/c^2$ --- thus the tidal force for BHs decreases as mass increases.  In stellar-mass systems, stars are observed to reach the tidal radius \cite[in this context, referred to as the Hill radius, or Roche limit;][]{fra85} persistently as opposed to transiently, and are observed as X-ray binaries \citep{pod02}.  While transient, tidal-disruption events (TDEs) can occur between stars and stellar-mass or intermediate-mass BHs in dense environments, such as globular clusters \citep[e.g.][]{ruiz09}, the present study focuses on extreme mass-ratio events in which stars are disrupted by the central, super-massive BHs (SMBHs) at the centres of galaxies.

By requiring the tidal radius to be larger than $\rs$, and considering a main-sequence star under the approximation that $\rstar \propto \mstar$, we obtain the maximum BH mass capable of observably disrupting a solar-type star \citep[e.g.][]{koch94},
	\begin{equation}
	\mbh \leq \mstar \scale[3/2]{\rsol}{2G\msol/c^2} \approx 10^8~\msol.
	\end{equation}

Stars cannot form within the tidal radius, thus for these disruptions to occur, stars must be scattered dynamically from farther away.  Models of two- and three-body encounters suggest that the feeding rate of stars to within $\rt$ could be roughly $10^{-6}$ to $10^{-4} \tr{ yr}^{-1}$ \citep[e.g.][]{mago99,wang04,mac12} for a typical luminosity galaxy (i.e.~$\sim L^{\star}$).

In the first analyses of TDEs, it was shown that these events could be characterized by a mass accretion rate which scales in time as roughly, $\dot{M} \propto t^{-5/3}$ \citep{rees88,phin89,evans89}.  The scaling can be derived by considering each element of the star ballistically, and decomposing the mass return rate into the distribution of binding energies as a function of mass, $d\epsilon/dM$, which is assumed to be constant, and the orbital period, which is given by Kepler's law, as a function of binding energy,
	\begin{equation}
	\frac{dM}{dt} = \frac{dM}{d\epsilon} \frac{d\epsilon}{dt} \approx \frac{1}{3} \frac{M}{\tm} \scale[-5/3]{t}{\tm}.
	\end{equation}
This relationship is generally expressed in terms of the characteristic, minimum-return timescale $\tm$, which is roughly the time at which the most-bound material will return to a pericentre $\rp$.  While simulations suggest that this is a good approximation for stars with relatively uniform densities \citep[e.g.][]{evans89}, \citet{gui13} have demonstrated that more accurate, centrally-condensed, stellar structures have smaller tidal radii, and give rise to different power-law accretion rates, typically in the range of $t^{-4/3}$ to $t^{-5/3}$, but partial disruptions can lead to a steeper time-dependence, $t^{-2.2}$.
 
Many studies assume that the mass return-rate can be taken as a proxy for the accretion rate onto the BH \citep[e.g.][]{vanv11b,guil14}, and further that this accretion rate will then be traced by the resulting light curve (e.g.~\citealt{komo99,geza08}; see however, \citealt{deco12}).  Post-disruption, however, the stellar debris is still in orbit, albeit with high eccentricities on the order of $e \gtrsim 0.9$.  For this material to be effectively accreted by the central BH, its orbital energy has to first be dissipated for circularization.  Recent investigations have shown this to be non-trivial \citep[e.g.][]{mac13}, and the underlying physical processes are actively being studied.

% ==== Introduction: Swift1644+57 =====
\subsection{Swift J1644+57}
\label{sec:sw}
Numerous candidate tidal-disruption events have been observed \citep[e.g.][]{komo99,geza08,vanv11a,cenk12,cho13}.
The Swift BAT initially triggered on Swift J1644+57 (\sw)~in March of 2011, and showed repeated X-ray flares with durations on the order of $10^3$ seconds, followed by a gradual decline over the next $10^7$ seconds.  The long term evolution in luminosity was observed to be consistent with the $t^{-5/3}$ power-law expected of a TDE \citep*[e.g.][]{blo11,bur11,lev11,zau11,tchek14}.  Additionally, the source was observed to have angular coincidence with the nucleus of a galaxy at redshift $z\approx0.35$ \citep{lev11}.  Finally, the observed emission (believed to be produced by a `jet' --- discussed shortly) was recently observed to `shut-off' \citep{zau13} --- consistent with a BH accretion-disc state transition.  Together, these lines of evidence suggest that \sw~is most likely a TDE from a SMBH.  While alternative theories have been proposed \citep[e.g.][]{kro11,qua12}, they generally suffer from systematic issues, for example in explaining the observed duration of the signal and coincidence with the centre of the host galaxy, in addition to the jet shutoff.

No emission, flaring or otherwise, has previously been observed from the galactic-nucleus associated with \sw---inferred to contain a central BH of mass $\approx 10^5\text{--}10^6 \, \msol$, discussed in detail in Appendix~\ref{sec:app_bhmass}.  The BH mass is important to constrain the magnetic flux \textit{available} to power the observed \sw~outburst.  Similarly, understanding the true luminosity of the event (discussed here) is critical to inferring the magnetic flux \textit{required} to match observations.  The interplay between these fluxes (\textit{available} vs.~\textit{required}) forms the foundation of this study.

The isotropic equivalent luminosity of \sw~in the X-ray was roughly $L_{iso} \approx \lumfe \, 10^{48} \tr{ erg s}^{-1}$, and the isotropic equivalent fluence, $10^{53}$---$10^{54}$ erg.  Observations of a radio afterglow associated with the event suggest that the ejected material producing the emission was at least mildly-relativistic, with a bulk Lorentz factor $\gamma \approx 2 - 20$ \citep{blo11,ber12}.  The beaming factor $f \equiv \frac{1}{2}(1 - \cos(\theta_j))$ describes the fraction of the sky covered by a jet with half-opening angle $\theta_j$.  The presence of a radio-afterglow associated with \sw~also argues for a collimated, relativistic outflow \citep{gian11,zau11,met12,ber12,Wiersema+12,zau13}. Based on detailed modeling of the synchrotron radio emission from \sw, specifically the inferred break frequencies and the flux measured at those values, \citet{met12} find $f \sim 1 - 5\E{-3}$.  With a beaming factor $f = \bfac \cdot 10^{-3}$, the energy emitted by \sw~is closer to $10^{51}$ erg, consistent with a solar-mass reservoir of energy.  For a BH of mass $\mbh = \mbhfive \mfive$ (this fiducial BH mass is discussed in Appendix~\ref{sec:app_bhmass}), the peak luminosity is closer to,
	\begin{equation}
	\begin{split}
	L_{p} & = 10^{45} \tr{ erg s}^{-1} \, \lumfe \, \bfac, \\
		  &	\approx 80 \tr{ L}_\tr{Edd} \eqsp \mbhfive^{-1} \, \lumfe \, \bfac .
	\end{split}
	\end{equation}

\sw~is unique in having been (to our knowledge) quiescent in the recent past, and yet rapidly producing a jetted, relativistic outflow with emission likely-above the source's Eddington luminosity.  The additional novelty of observing a jet shutoff \citep[and perhaps a re-emission, in the future; see,][]{tchek14}, presents an incredible opportunity to understand the physics underlying jetted systems in general.  In this study, we use \sw~as a laboratory for studying jet production in an environment with relatively well-constrained initial conditions.

% ==== Introduction: Sw Gone MAD =====
\subsection{Gone MAD}
The \sw~event was quite different from the characteristic, smoothly-evolving, optical/UV transient expected from a tidal disruption.  The peak flux and power-law decline of the observed light curve is consistent with TDE expectations, along with the overall energy scale.  The rapid onset, short variability timescale, initial flaring, and the observed signal being in the X-ray, however, all challenge the standard scenario. 

Accreting BHs are believed to power relativistic, jetted outflows which extract spin energy from the disc and BH magnetically. This process is known as the Blandford-Znajek (BZ) mechanism \citep[][]{bla77,tch11} and requires a large-scale magnetic field, pushed into the BH by accretion, to power an outflow. The BZ luminosity can be expressed as,
	\begin{equation}
	\label{eq:bz-lum}
	L = 4 \pi \rs^2 \, c \, \xi_\trt{BZ} \left( \frac{B^2}{8 \pi}\right) = \frac{c \, \xi_\trt{BZ} }{2 \pi^2 \rs^2} \Phi^2,
	\end{equation}
where $\xi_\trt{BZ} \approx 0.1 a/[1+(1-a^2)^{1/2}]$ is an efficiency factor related to the black-hole spin $a$ \citep{tch10a}, $c$ is the speed of light, and $\Phi$ is the magnetic flux threading the BH.  The observed, peak luminosity can then be translated into a minimum total flux threading the BH of,
	\begin{equation}
	\label{eq:flux_jet}
	\Phi_\tr{jet} \gtrsim 8\E{28} \tr{ G cm}^2 \eqsp \mbhfive \, \lumfe^{1/2} \, \bfac^{1/2}.
	\end{equation}

\citet*{tchek14} (hereafter \tchek) described a model that is capable of explaining the unexpected aspects of \sw.  
During a TDE the angular momentum axis of the debris and the transient accretion disc (given by the angular momentum of the stellar orbit) is expected to be misaligned relative to the BH spin axis.  During the initial phases of outburst, when the mass accretion rate is highest, the BH magnetic field is sub-dominant, and the jet is launched along the angular momentum axis of the transient disc.  As the accretion rate declines---and, perhaps, BH magnetic flux builds up---the BH magnetic flux becomes dynamically-important, and the accretion flow becomes a \emph{magnetically arrested disc} \citep[MAD,][]{nar03, tch11}.
In MADs, the magnetic field is so strong that it aligns the jet axis with BH spin axis \citep*{2013Sci...339...49M}.  If the BH spin axis points at us, the jet's beamed, high-energy emission becomes visible.  However, the process of alignment is not clean: as the magnetic flux works to realign the jet, the jet punches holes in the disc, and wobbles chaotically, producing a flare every time it passes in front of our line of sight.

One of the requisite characteristics of BZ-powered jets is a large-scale magnetic flux \citep{mcki04, beck08, mcki12}. According to Eq.~\ref{eq:flux_jet}, any BZ-powered jet requires a very strong BH magnetic flux to power a luminous outburst like \sw.   It is particularly puzzling that a completely quiescent system was able to generate such a powerful relativistic jet.  In particular, as we show in \tsec{sec:swift1644}, the required magnetic flux (Eq.~\ref{eq:flux_jet}) is 2--3 orders of magnitude larger than what could have been threading the quiescent BH, and 3--5 orders larger than that of a main-sequence star.  Where could such an enormous magnetic flux come from?  The typical mechanism for amplifying magnetic field in accretion disc systems is the magnetorotational instability \citep[MRI,][]{bal91}.  The MRI, however, produces a tangled field with no net flux.  Whether local flux-excesses from the MRI can produce jets has yet to be quantitatively constrained, but it seems unlikely that such a process could maintain a steady jet like \sw~for the observed periods of time.  A dynamo mechanism is also unlikely to suffice on timescales as short as a TDE -- comparable to the minimum orbital period of debris.
\tchek~postulate the existence of a sufficient flux threading the BH and suggest that it could have come from a pre-existing, ``fossil'' accretion disc. However, the mechanism by which this flux could be delivered to the BH remained unclear.

In this work, we show that the interaction of the debris stream with the fossil disc is unstable to surface instabilities. These instabilities lead to effective mixing of the fossil magnetic flux into the stream, which brings it toward the BH.
In \tsec{sec:problem_setup} we describe the inferred parameters of the \sw~event, namely those of the encounter, quiescent system, and resulting debris stream.  We simplify this setup in \tsec{sec:toy_model} into a `toy model' for further analysis.  In \tsec{sec:results}, we describe our numerical simulations and show that the interface between the stellar debris stream and the fossil disc is dynamically unstable (\tsec{sec:instab}) to instabilities that are effective at capturing the magnetic flux from the disc, and carrying it to the BH (\tsec{sec:flux-transport}).  Our conclusions and a discussion of the ramifications of our results are presented in \tsec{sec:conc}.  Finally, some additional details of our numeric and analytic procedures are included in appendices \tsec{sec:app_sims} and \tsec{sec:app_global} respectively.

% ===============================================================================
% ============================== PROBLEM SETUP ==================================
\section{Problem Setup}
\label{sec:problem_setup}

% ==== Problem Setup: Swift 1644+57 =====
\subsection{Swift 1644+57}
\label{sec:swift1644}
The tidal-disruption event \sw~did not conform to any existing model of a tidal disruption \citep[or other transient phenomenon; see however,][]{gian11}, making parameter estimation difficult.  Recent efforts to model the event contain significant variance in the inferred physical parameters.  Nonetheless, the characteristic timescale and energetics of the event suggest the disruption of a roughly solar-mass star by an SMBH.  In the following sections, we attempt to constrain the parameters of the disruption.

% ====== STELLAR ENCOUNTER AND DEBRIS =======
\subsubsection{Stellar Encounter and Subsequent Orbits}
\label{sec:stellar}
A tidal disruption begins as a simple problem.  The system is, for practical purposes, defined entirely by the star and BH's masses, $\mstar = \mstarz \, \msol$ and $\mbh = \mbhfive \, \mfive $.  The impact parameter, $\beta \equiv \rt/\rp$, describes the effective depth of the encounter as the ratio of the tidal radius $\rt$, to the pericentre radius $\rp$.  Recent work has found, however, that the energy distribution in the stellar material is effectively frozen in when the star passes the tidal radius, regardless of its eventual pericentre distance \citep{gui13,sto13}.  This does rely on the assumption that the star's pericentre distance is just within the tidal radius: i.e.~$\beta > 1$, while still of order unity.

Given a pair of masses, the tidal radius is,
	\begin{equation}
	\begin{split}
	\rt  	& = \rstar \scale[1/3]{\mbh}{\mstar}, \\
			& = 3.2\E{12} \tr{ cm} \, \mstarz^{2/3} \, \mbhfive^{1/3}, \\
			& = 110 \, \rs \eqsp \mstarz^{2/3} \, \mbhfive^{-2/3},
	\end{split}
	\end{equation}
where $\rs$ is the Schwarzschild radius, and here and hence-forth we make the approximation that $\rstar = \rsol {\mstar}/{\msol}$.  The minimum return time\footnote{Note that in \tchek, the given $\tm$ included the redshift correction for \sw, $z\sim0.35$, i.e.~$\tm(1+z)\approx1.5\E{6}$ s.  Here we use rest-frame quantities.} can then be defined as \citep[][with $\rp \rightarrow \rt$]{evans89},
	\begin{equation}
	\begin{split}
	\tm 	& =  \frac{\pi}{\sqrt{2}} \scale[3/2]{\rt}{R_\star} \scale[1/2]{\rt}{G \mbh}, \\
			& = 1.1\E{6} \tr{ s} \eqsp \mbhfive^{1/2}  \,\mstarz^{1/2}.
	\end{split}
	\end{equation}
The accretion rate peaks near $1.5 \, \tm$ \citep{evans89,sto13}, and is described by,
	\begin{equation}
	\begin{split}
	\label{eq:tde-mdot}
	\dot{M} & = 5.9\E{26} \tr{ g s}^{-1} \eqsp \ttime^{-5/3} \, \mstarz^{1/2} \, \mbhfive^{-1/2}, \\
			& = 4.2\E{4} \,\, \dot{M}_\trt{Edd} \eqsp \ttime^{-5/3} \, \mstarz^{1/2} \, \mbhfive^{-3/2},
	\end{split}
	\end{equation}
where the time is parametrized as, $t = \ttime \cdot \tm$, and $\dot{M}_\trt{Edd}$ is the Eddington accretion rate.  Immediately after the star is disrupted, and before the debris can be accreted, the stellar material orbits with a semi-major axis $\major$, given directly by the orbital return time as,
	\begin{equation}
	\begin{split}
	\label{eq:major-axis}
	\major 	&	= \left( G \mbh \right)^{1/3} \scale[2/3]{t}{2\pi}, \\
			&	= 2500 \, \rs \eqsp \ttime^{2/3} \, \mstarz^{1/3} \, \mbhfive^{-1/3},
	\end{split}
	\end{equation}
which is determined solely by the binding energy of stellar material at tidal-radius passage.  The semi-minor axis $\minor$, determined by the specific angular momentum, can be expressed as,
	\begin{equation}
	\begin{split}
	\label{eq:minor-axis}
	\minor	&	= \left( 2 \rstar \right)^{1/2} \left( G \mbh \mstar \right)^{1/6} \scale[1/3]{t}{2\pi}, \\
			&	= 740 \, \rs \eqsp  \ttime^{1/3} \, \mstarz^{1/2} \, \mbhfive^{-1/2}.
	\end{split}
	\end{equation}
For the fiducial parameters, the tightest bound material starts at an eccentricity of about $e\sim0.96$, and increases to $0.99$ by $10 \, \tm$.

% ====== DEBRIS STREAM =======
\subsubsection{Debris Stream}
\label{sec:debris}
Assuming that self-gravity is unimportant \footnote{This subject, along with the internal structure of the debris stream, is discussed in detail in Appendix~\ref{sec:app_stabil}.}, the solid-angle subtended by the debris stream, as viewed by the central SMBH, is approximately constant in time \citep{str09},
	\begin{equation}
	\begin{split}
	\Delta \Omega	&	\approx \sqrt{48} \scale[3/2]{\rstar}{\rp} , \\
					&	= 0.02 \eqsp \mstarz^{1/2} \, \mbhfive^{-1/2}.
	\end{split}
	\end{equation}
Using this solid angle, the debris density can be related to the accretion rate as,
	\begin{equation}
	\label{eq:mdot-dens}
	\dot{M} = \strdens \frac{\Delta \Omega}{4\pi} R^2 v_s \, ,
	\end{equation}
where $R = r \, \rs$ is the radial coordinate from the SMBH, and $\strdens$ and $v_{str}$ are the debris-stream density and velocity respectively.  Assuming a free-fall velocity, and the mass accretion rate from Eq.~\ref{eq:tde-mdot}, the density is then approximately,
	\begin{equation}
	\begin{split}
	\label{eq:tde-dens}
	\strdens = 1.3\E{-2} \, \frac{\tr{g}}{\tr{cm}^{3}} \eqsp \rrad^{-3/2} \, \ttime^{-5/3} \, \mbhfive^{-2}.
	\end{split}
	\end{equation}
Throughout our analysis we model the stream as a uniform density cylinder, with an effective radius inferred from Eq.~\ref{eq:mdot-dens}, i.e.,
	\begin{equation}
	\begin{split}
	R_{str} &	\approx \left( \frac{\Delta\Omega}{4\pi}\frac{R^2}{\pi} \right)^{1/2}, \\
			&	= 7.0\E{8} \tr{ cm} \eqsp \rrad \, \mstarz^{1/4} \, \mbhfive^{3/4}.
	\end{split}
	\end{equation}	
	
During a tidal encounter, the star is stretched nearly-radially from the BH which leads to its disruption.  At the same time, it is compressed into the plane of the orbit.  The maximum compression can be considered analytically \citep[e.g.][]{lumi86,sto13}, or calculated numerically.  As a reference, \citet{gui09} find a compression $p_{\star,max} \approx 1.3\E{18} \tr{ erg cm}^{-3}$ when the peak density is $\rho_{\star,max} \approx 310 \tr{ g cm}^{-3}$ for a deep ($\beta = 7$) encounter.  After peak compression the star rebounds, and expands along the orbit as it evolves into a debris stream.

We assume that after pericentre passage the star expands adiabatically.  Using $p_{\star,max}$, and the density from Eq.~\ref{eq:tde-dens}, 
	\begin{equation}
	\label{eq:str-pressure}
	\begin{split}
	p_{str}	&	= p_{\star,max} \scale[\gamma]{\strdens}{\rho_{\star,max}}, \\
			&	=  6.6\E{10} \, \frac{\tr{erg}}{\tr{cm}^{3}} \eqsp \rrad^{-5/2} \, \ttime^{-25/9} \, \mbhfive^{-10/3},
	\end{split}
	\end{equation}
for an adiabatic index $\gamma = 5/3$.  For a more modest impact parameter---as expected for a TDE like \sw---the stream pressure will be lower.  For a $\beta \approx 1$ impact, the overall change in volume will be of order unity, and additional heating effects \citep[e.g.][]{cart83,bras08} should be negligible (see however, Appendix~\ref{sec:app_stabil}); we thus consider the value in Eq.~\ref{eq:str-pressure} as an upper limit.  

The debris stream itself can be expected to have a magnetic flux negligible compared to that needed to power the observed X-ray flare.  The required flux is on the order of $10^{29} \textrm{ G cm}^2$ (a field strength of about $10^7$ G in our fiducial model; see Eq.~\ref{eq:flux_jet}), while stellar magnetic fluxes could\footnote{Assuming ordered fields and large filling factors.} reach up to $10^{23} \textrm{ G cm}^2$ (fields of 10 -- 100 G) and perhaps, occasionally $10^{25} \textrm{ G cm}^2$ \citep[e.g.][]{bran05,hubr13}.  Because the stellar debris is spread over length scales 100's -- 1000's of times its initial stellar radius (see Eqs.~\ref{eq:major-axis}, \ref{eq:minor-axis}), any initial fields will be drastically diluted due to flux freezing.  Assuming an initially isotropic magnetic field, with a typical, stellar strength, ${\bstar}_\trt{init} = \bstarz \, \cdot 1 \tr{ G}$, we can estimate the resulting magnetic field strength as,
	\begin{equation}
	\label{eq:b-str}
	\begin{split}
	\bstr 	&	\approx \bstarz \scale[2]{\rstar}{\rstr}, \\
			& 	= 10^4 \tr{ G} \eqsp \bstarz \, \mstarz^{3/2} \, \mbhfive^{-3/2} \rrad^{-2}.
	\end{split}
	\end{equation}
Note that in our convention the radial coordinate is normalized to the Schwarzschild radius, i.e.~$R = r \, \rs$. The field strength in Eq.~\ref{eq:b-str} falls off rapidly with distance, and thus also with time.  At apocentre, $R \approx \major \approx 2000 \, \rs$, the field strength drops to $\bstr \approx 10^{-3} \textrm{ G}$.  This field can be assumed to orient along the direction of maximum expansion---along the stream axis ($\hat{x}$)---as this is the field-component least diluted by flux-freezing during expansion.  The field component out of the plane of the orbit ($\hat{z}$) will be diminished the most.

% ====== QUIESCENT BLACK HOLE =======
\subsubsection{Quiescent Black Hole}
\label{sec:qui-bh}
Archival observation of the host galaxy's area on the sky provide X-ray and radio upper-limits of $L_\trt{x,pre} \lesssim 1.7\E{44} \tr{ erg s}^{-1}$ and $F_\trt{rad,pre} \lesssim 0.3 \tr{ mJy}$ respectively, on the luminosity of the pre-disruption SMBH \citep{blo11}.  As no previous activity has been observed in this system it is impossible to definitively constrain the pre-disruption state of the SMBH or its galactic centre.

Observations following up the Swift detection showed the presence of a radio counterpart \citep{ber12}, which continued to rise over many days.  These radio light curves are well fit by an afterglow model \citep{met12}, in which the transient jet shocks and decelerates into the ambient medium.  These fits yield an indirect measurement of the ambient density profile around the transient.  The densities inferred are consistent with a Bondi profile \citep{ber12} of number density,
	\begin{equation}
	n = 2.0 \tr{ cm}^{-3} \scale[-3/2]{R}{0.1 \tr{ pc}}.
	\end{equation}
The density at these distances reflects the conditions of the `quiescent' state of the galaxy's centre, preceding the TDE.  In particular, we can constrain the quiescent accretion rate.  Assuming a steady-state, isotropic inflow,
	\begin{equation}
	\label{eq:acc}
	\dot{M}_q = 4 \pi R^2 \, v \, m_p \, n,
	\end{equation}
for a flow of speed $v$ and a number density $n$.  An upper limit to the accretion rate is given by assuming the flow is in free-fall, with $v = c \left( \rs/R \right)^{1/2}$.
Using the inferred Bondi density profile, the quiescent accretion rate can be constrained to,
	\begin{equation}
	\begin{split}
	\label{eq:qui-mdot}
	\dot{M}_q 	&	\lesssim 3.7\E{19}   \textrm{ g s}^{-1}  \eqsp  \mbhfive^{1/2}, \\
				&	\lesssim 6\E{-7}  \eqsp \msol \textrm{ yr}^{-1}  \mbhfive^{1/2}, \\
				&	\lesssim 2.6\E{-4} \eqsp \dot{M}_\textrm{Edd} \, \mbhfive^{-1/2}, 
	\end{split}
	\end{equation}
where we define $\dot M_{\rm Edd} = 10 \, L_{\rm Edd}/c^2$.

The maximum strength magnetic field which could be present in the quiescent system is constrained to equipartition with the mass inflow ($B^2_{q}/8 \pi \approx \rho \, v_\trt{ff}^2$), i.e.,
	\begin{equation}
	\label{eq:equi-b}
	B_q^2 = \frac{\dot{M} c}{R^2_s} \left( \frac{\rs}{R} \right)^{5/2},
	\end{equation}
which, for the accretion rate given in Eq.~\ref{eq:qui-mdot}, evaluates to,
	\begin{equation}
	\label{eq:mag}
	B_q 	= 3.6\E{4} \tr{ G} \eqsp \rrad^{-5/4} \, \mbhfive^{-3/4}.
	\end{equation}
Expressed in terms of the magnetic flux threading the BH, this is,
	\begin{equation}
	\begin{split}
	\Phi_\bullet 	& \approx \pi \rs^2 \,\cdot\, B_q(\rs), \\
					& = 10^{26} \textrm{ G cm}^2 \eqsp \mbhfive^{5/4}.
	\end{split}
	\end{equation}

In Sec.~\ref{sec:sw} we showed that the magnetic flux required to power \sw, was $\Phi_\tr{jet} \approx 8\E{28} \tr{ G}$ (Eq.~\ref{eq:flux_jet}).  The maximum quiescent field would thus need to be amplified, or otherwise augmented, by almost three orders of magnitude to explain the observed jet luminosity.  While the flux threading the quiescent BH is insufficient, there could easily be sufficient flux spread throughout the disc.  The radius within which sufficient magnetic field exists can be calculated by integrating Eq.~\ref{eq:mag} and setting the result equal to the required field.  This yields a radius,

	\begin{equation}
	\begin{split}
	R_\trt{enc} 	\approx 4.1\E{3} \, \rs \eqsp \bfac^{2/3} \,\, \lumfe^{2/3} \, \mbhfive^{-1/3}, 
	\end{split}
	\end{equation}
which is within a factor of $2$ of the debris stream's initial semi-major axis.  Thus, while a sufficient magnetic flux does not initially thread the BH, it is easily present within the region of the TDE.  This reservoir motivates our flux-capture model, in which the stellar debris stream ``captures'' the magnetic field in the ``fossil'' disc, and carries it to the BH.

% ====== QUIESCENT DISC =======
\subsubsection{Quiescent Disc}
\label{sec:qui-disk}

To understand interactions between the stellar debris stream and the quiescent system, we use an advection dominated accretion flow (ADAF) model \citep{nar94} to derive realistic parameters matched to the inferred accretion rate.

The sound speed for an ADAF is given by \citep{nar98}, 
	\begin{equation}
	\label{eq:adaf-cs}
	c_s^2 = \frac{2(5+2\epsilon')}{9\alpha^2} \cdot g(\alpha, \epsilon') \cdot \frac{G\mbh}{R},
	\end{equation}
where the functions $g$ and $\epsilon'$ are,
	\begin{equation}
	\begin{split}
	g(\alpha,\epsilon') & \equiv \sqrt{1 + \frac{18\alpha^2}{(5+2\epsilon')^2}} - 1, \\
	\epsilon' 			& \equiv \frac{1}{f_{re}}\left( \frac{5/3 - \gamma}{\gamma - 1} \right).
	\end{split}
	\end{equation}
This model assumes a viscosity parameter $\alpha$ \citep{sha73}, radiative efficiency $f_{re}$, and uses an adiabatic index (ratio of specific heats) $\gamma$.  Because the magnetic field can be dynamically important, $\gamma$ is not simply based on the gas properties.  For a total pressure composed of thermal and magnetic components, $p=p_g + p_B$, we can define their relative significance using $\pbeta$ as $p_g = \pbeta \cdot p$.  Then the effective adiabatic index is,
	\begin{equation}
	\gamma = \left(\frac{32-24\pbeta-3\pbeta^2}{24-21\pbeta}\right).
	\end{equation}
An equipartition magnetic field with $\pbeta \sim 0.5$ yields an index of $\gamma \sim 1.43$.  Assuming, then, that the disc is radiatively inefficient, $f\sim 1.0$, this yields, $\epsilon' \sim 0.56$, for a sound speed,
	\begin{equation}
	c_s = 1.2\E{10} \tr{ cm s}^{-1} \eqsp \rrad^{-1/2}.
	\end{equation}
Finally, the pressure of the gas can be determined as $P=\rho c_s^2$, i.e.
	\begin{equation}
	P = 1.6\E{7} \tr{ erg cm}^{-3} \eqsp \rrad^{-5/2} \, \mbhfive^{-3/2}.
	\end{equation}

% ====== TOY MODEL =======
\subsection{Toy Model}
\label{sec:toy_model}
There are only a few, simple initial parameters which define the general dynamics of a tidal-disruption.  Nonetheless, the resulting dynamics are rich and complex, especially in their geometry and evolution in time.  In our analysis, we can simplify the global picture into a zoomed-in toy model which captures the relevant physics.  Here we outline our expectations of those most-relevant phenomena, and describe our model which incorporates them.

As the most-bound stellar material returns to the BH, it defines the inner-most ellipse of the debris orbits.  All of the following material returns from larger ellipses, but during the circularization process, all orbits entering $R_{\rm circ}$ will gradually shrink towards the BH.  The quiescent-disc material which was inside $R_{\rm circ}$, and in the plane, will presumably be dragged along with the debris into the BH, along with the magnetic flux in this region. However, since $R_{\rm circ}\sim 100 \, \rs$, this magnetic flux is insufficient.

The rotation of the quiescent disc into the stream brings a sizeable fraction of the total disc into contact with the debris stream out to its apocentre at $\sim 2 \major$.  This presents a significant increase to the available field, and importantly includes a region with a sufficient amount of magnetic flux to power the observed \sw~transient (\tsec{sec:qui-bh}).
We would like to understand if the debris stream is able to `capture' flux from the disc material it encounters.  In effect, the stream is cold, high-density, unmagnetized material passing through a hot, low-density, magnetized material.  For the flux to be captured, some sort of mixing must occur.

The interaction between the disc and stream can be susceptible to both Rayleigh-Taylor (RT) and Kelvin-Helmholz (KH) instabilities.  In the context of KH instability, the stream is moving almost radially, while the disc is rotating nearly tangentially, thus producing a significant velocity shear between the two.  In the context of RT instability, the light, magnetized disc decelerates into the dense debris.

Any numerical model employed to demonstrate mixing must be capable of capturing these effects.  At the same time, simulating the entire problem requires an enormous range of size scales, from the BH itself, to the apocentre of the debris stream.  Resolving small-scale turbulence in the debris stream, at the same time, is seemingly intractable with current computational capabilities.

	% FIGURE: Toy Model Schematic (1)
	\begin{figure}
	\begin{center}
	\includegraphics[width=0.95\columnwidth,height=0.75\columnwidth]{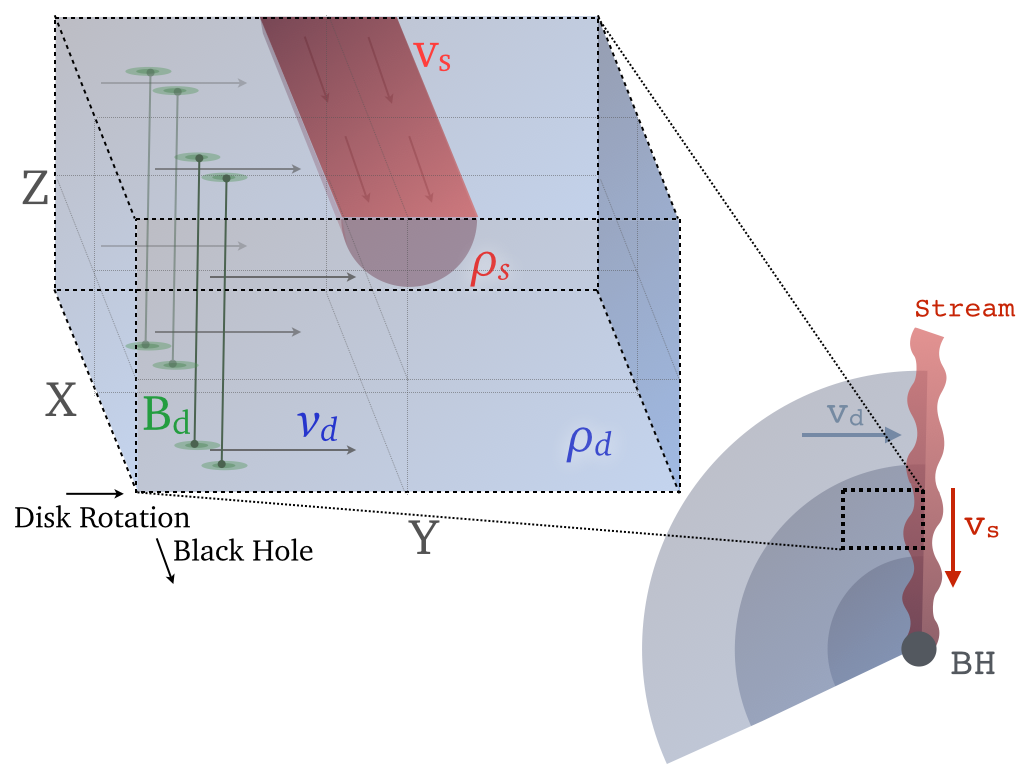}
	\end{center}
	\caption{Schematic of the `toy model' used for calculating magnetic flux transport.  The model focuses on a region of the disc--stream interface, at a distance away from the BH such that the disc and stream can be assumed stationary over the relevant timescales.  The stream is initialized with a velocity $v_s \, \hat{x}$ and density $\strdens$ which is significantly larger than the disc density $\rho_d$.  The disc moves orthogonally with velocity $v_d \, \hat{y}$, carrying in an initially uniform magnetic field $B_d \, \hat{z}$.}
	\label{fig:toy}
	\end{figure}

To make progress, we consider a simplified toy model that allows us to capture the physics of small-scale instabilities. It is illustrated schematically in Fig.~\ref{fig:toy}.  We consider a small region of the disc, far from the BH, and model the stellar debris as a cylindrical stream of density $\strdens$, injected into the disc.  This stream moves radially towards the BH with a velocity $v=v_s \hat{x}$.  As the internal shearing of the disc is unimportant on the scales of interest, we consider the disc to be in uniform motion with a velocity $v=v_d \hat{y}$.  Far from the BH, stratification and gravity are not important for the evolution of purely hydrodynamic instabilities---like RT and KH.  Thus, we consider a uniform disc density $\rho_d$, with a uniform magnetic field, $B=B_d \hat{z}$, and no gravity.

We assume that the magnetic field is perpendicular to the debris' orbital plane. This is a reasonable approximation for the following reasons.  If the fossil magnetic field is misaligned, tangled, or even dominantly toroidal, the effect of flux draping is to amplify the $\hat{z}$ component \citep{vik01}.  Field lines that do not cross the orbital midplane go around the stream, above and below it.  Vertically oriented components on the other hand will be draped, and amplified.

The physical scenario involves a drastic density contrast between the stream and the disc, $\strdens/\rho_d \approx 10^{11} \eqsp \mbhfive^{-1/2} \, \ttime^{-5/3}$.  The stream's velocity is supersonic relative to the disc sound speed.  And additionally, because the stream is cold, both the stream and disc velocities are highly supersonic relative to the stream's sound speed.  As the disc rams against the debris stream, the magnetic field lines drape around the stream, increasing their strength at the interface, similar to the `flux draping' problem studied in the context of galaxy clusters \citep[e.g.][]{vik01,lyu06,mar07,dur07}.  In the region where the field is amplified, the disc flow also stagnates --- producing a highly magnetic-energy dominated region.  The large density contrasts, supersonic flows, and magnetically-dominated regions make this a challenging problem numerically.  To make the problem tractable for magnetohydrodynamic (MHD) codes, we soften the input parameters.  In particular, we explore a more practical density contrast of $10^3$.
This density ratio, still much larger than unity, allows us to capture the important physics and scale it to realistic values.

% =============================================================================
% ================================= RESULTS ===================================
\section{Numerical Simulations and Results}
\label{sec:results}

% ====== MAGNETOHYDRODYNAMICS =======
\subsection{Magnetohydrodynamics}
\label{sec:magnetohydrodynamics}
We have performed numerical simulations using the Athena MHD code \citep[v4.2;][including stability improvements, see Appendix~\ref{sec:app_sims}]{sto08} of the ``toy'' problem described in \tsec{sec:toy_model}.  To first order, our toy model is a two dimensional problem --- translationally invariant along the stream axis.  Only the onset of turbulence breaks the symmetry and leads to magnetic flux transport.  Even the purely hydrodynamic (HD) situation, however, of an infinite cylinder embedded in a uniform flow field has no known, closed form solutions in the general, compressible case \citep[e.g.][]{cha61,roy86,gla88,wu91}.

	% FIGURE: 2D MHD - x-slices - Density, Magnetic and Pressure Fields (2)
	\begin{figure*}
	\includegraphics[width=1.0\textwidth,height=0.45\textwidth]{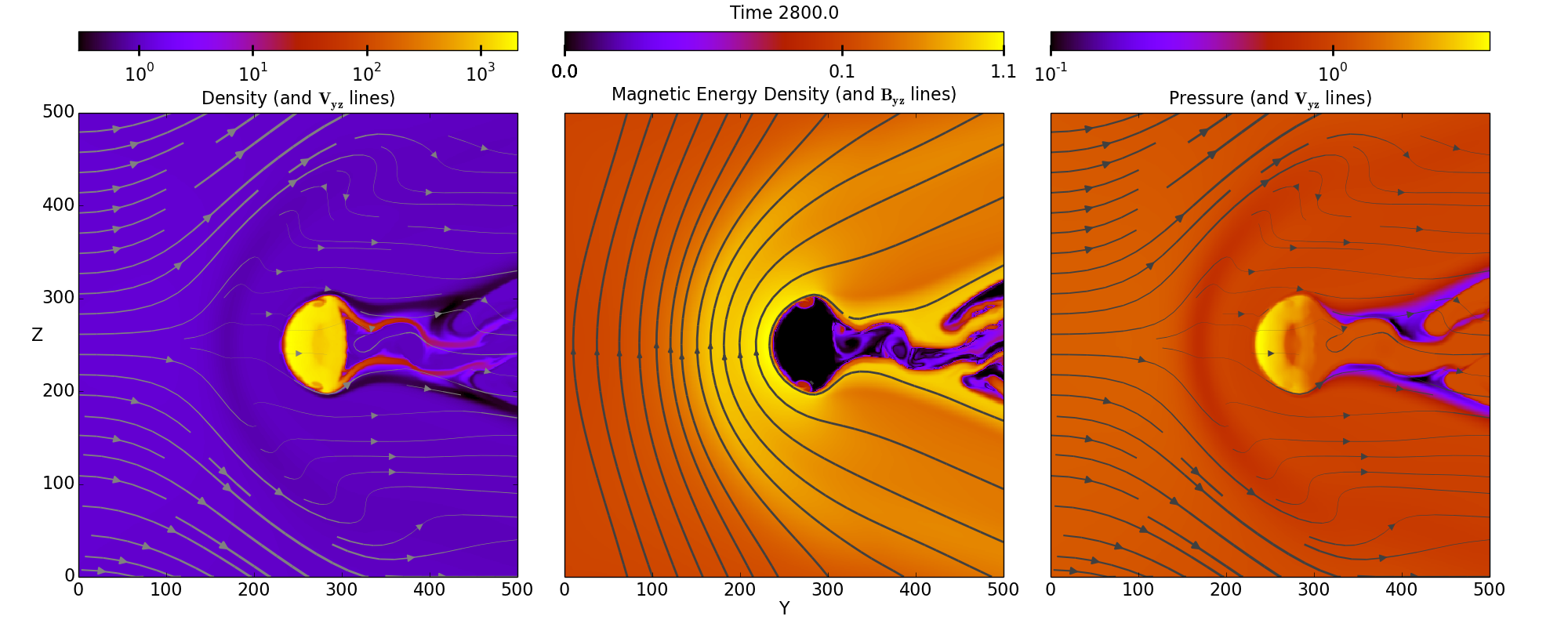}
	\caption{The three panels show, from left to right, density, magnetic energy density, and pressure in 2D MHD simulations with a resolution of $512\times512$ cells.  The velocity field is over-plotted with streamlines in the density and pressure (left and right) panels, with line-thickness proportional to the speed.  Magnetic field lines are plotted over the magnetic energy density (centre) panel.  The stream is moving with a velocity $v_s = 3\hat{x}$ (out of the page), and the disc with velocity $v_d = 0.5\hat{y}$.  Note that the magnetic energy density (centre panel) uses a broken, semi-logarithmic color scaling---linear between 0.0 and 0.1.  This snapshot is taken a little before 3 disc-crossing times.  The magnetically-dominated, stagnation region ahead of the stream, and turbulent wake behind it, are clearly apparent in all panels.  Behind the stagnation region, the velocity is significantly depressed along with the thermal pressure, while the magnetic field strength grows and drapes around the stream.  \movielink{http://youtu.be/5sciTO7j6M8} }
	\label{fig:2dmhd}
	\end{figure*}

	% FIGURE: 2D HD vs. MHD - v2 field (3)
	\begin{figure*}
	\includegraphics[width=1.0\textwidth,height=0.65\textwidth]{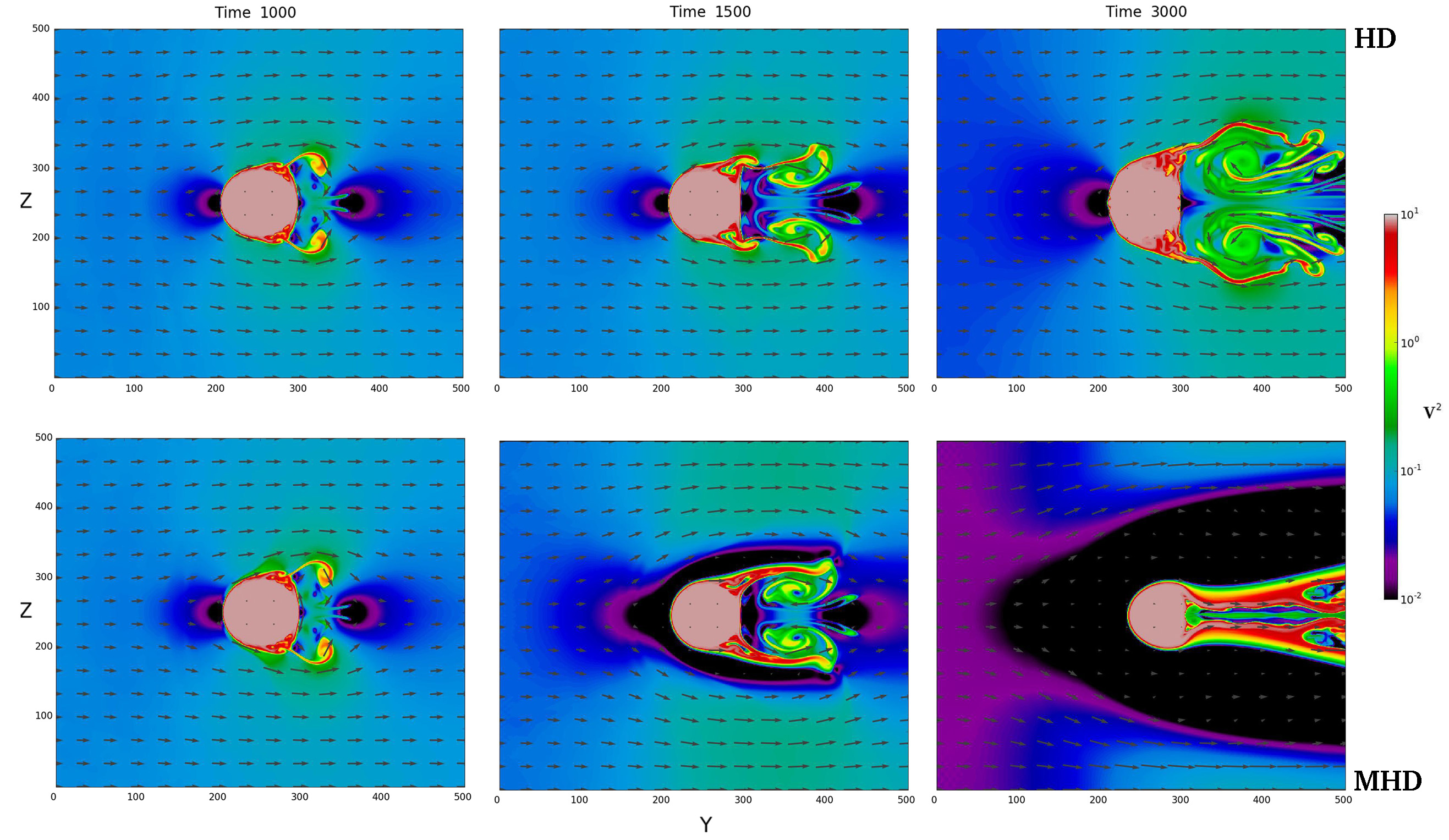}
	\caption{Comparison between 2D hydrodynamic (HD - top row) and magnetohydrodynamic (MHD - bottom row) simulations in 2D, at three different times.  The effects of magnetic field pile-up are pronounced, despite a background magnetic field with an energy density more than an order of magnitude below thermal.  A very large `stagnation region' is present in the shape of a bow-shock, which significantly increases the overall cross-section of interaction between the stream and disc.  The effect of an increased standoff distance at the interface also serves to separate the boundaries unstable to Kelvin-Helmholtz (directly at the stream) and Rayleigh-Taylor (ahead of the stagnation region) instabilities.}
	\label{fig:hd-vs-mhd}
	\end{figure*}
 
As illustrated schematically in Fig.~\ref{fig:toy}, we inject the stream along the $\hat{x}$ axis, and the disc in the $+\hat{y}$ direction.  The grid is $500$ simulation units along all axes, with the stream initialized to a cylinder with a radius of $50$ units, centred around the centre of the y-z--plane.  The stream is initialized with a velocity of $v_s = 3.0$, while the density ($\rho_d$) and pressure ($p_d$) of the disc are set to unity.  The stream velocity then corresponds to a Mach number $\mathcal{M}_s \approx 2.3$ relative to the disc sound speed (for an adiabatic equation-of-state with $\gamma = 5/3$).  The background disc is initialized to zero velocity and zero magnetic field for stability reasons, and additional material is injected from the boundary ($y=0$) with a velocity $v_d = 0.5$---a Mach number of $\mathcal{M}_d \approx 0.4$---and magnetic field $B = B_d \hat{z}$, such that the magnetic energy density $\varepsilon_\trt{B} = B^2/8\pi = 0.02$.

\subsubsection{2D Dynamics}
While the instabilities that we are interested in only exist in the full 3D simulations, the most basic properties of the evolution can be isolated and elucidated by 2D simulations of a cross-section of the disc and debris stream.  Such a cross-section is depicted in Figure~\ref{fig:2dmhd}, which shows the density, magnetic-field, and thermal pressure.  As the disc flows around the stream, a wake rapidly forms with strands of stream material detaching from the main body soon after.  Strong vortical regions are readily apparent.  The velocity in the y-z plane ($V_{yz}$) is significantly depressed within the magnetic-field draping region - in the shape of a bow-shock around the stream.

The disc is injected sub-sonically, and thus a hydrodynamic bow-shock does not form.  However, the magnetic fields, which drape around the stream, stall the inflowing material - causing a stagnation region.  Fig.~\ref{fig:hd-vs-mhd} shows the effects of magnetic field on the simulation by comparing the velocity field magnitude ($|V|^2$) in a purely hydrodynamic run (upper panels) to one with magnetic fields (lower panels), at three different times.  These 2D simulations have identical parameters except for the magnetic field strength.  Even though the magnetic field is injected with an energy density only $2\%$ of thermal ($16\%$ of the disc inflow ram pressure), the flow geometry is drastically altered.  Note that the interaction cross-section of the stream is significantly enhanced.

	% FIGURE: 2D MHD - midplane slices over time - Density, Vx, Bz (4)
	\begin{figure*}
	\includegraphics[width=1.0\textwidth,height=0.45\textwidth]{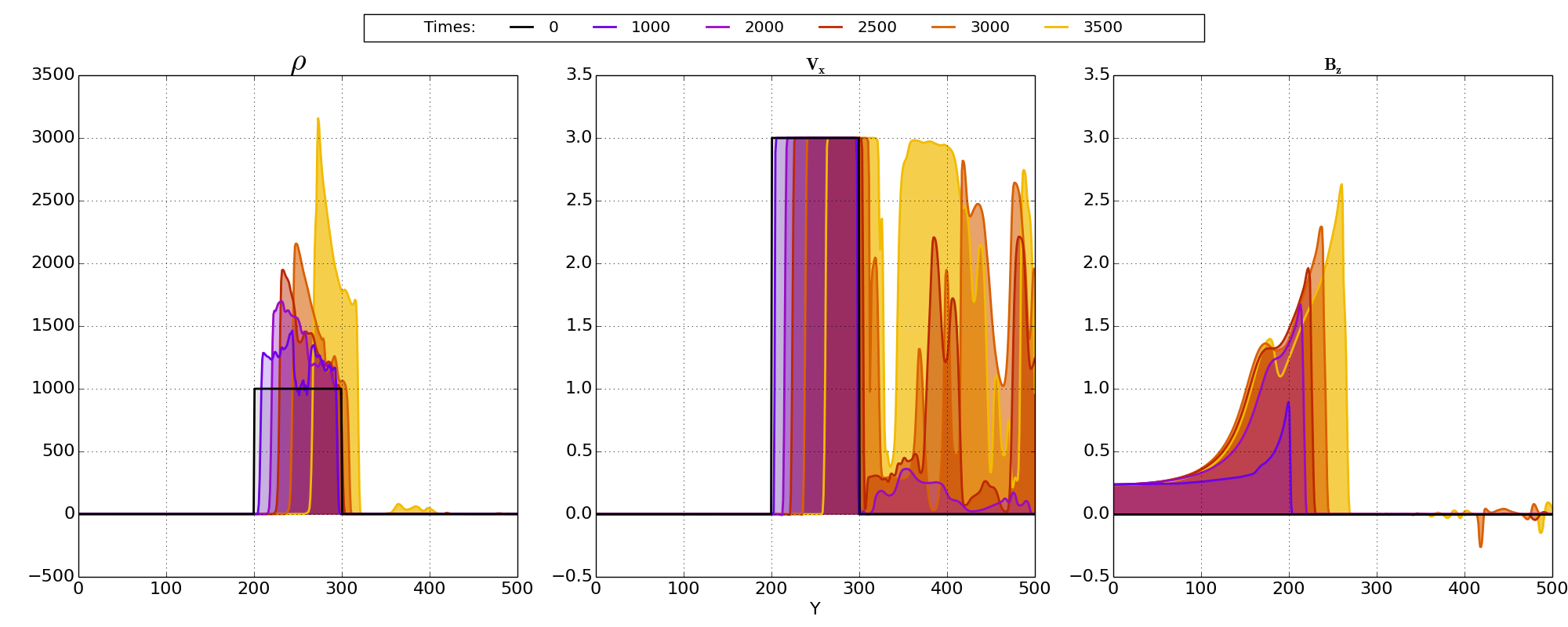}
	\caption{From left to right, panels show slices of density $\rho$, $x$-velocity $V_x$, and vertical magnetic field $B_z$ measured in the z-midplane of 2D MHD simulations.  The stream moves in the $\hat{x}$ direction (out of the page), and the disc along $\hat{y}$ (along the abscissa).  Each property is plotted at time intervals of 500 time units---about a disc-crossing time.  The first snapshot (0: red) corresponds to $t=0$, after which the density distribution (left panel) shows that the stream is both contracted over time and translated along the y-axis due ram pressure from the disc.  The velocity structure (centre panel) shows the oscillations in the wake of the stream pulling high-velocity material away.  The $z$-component of magnetic field ($B_z$, right panel) grows dramatically at the stream edge --- maintaining a roughly $1/r$ profile in strength and growing roughly linearly with time (see the discussion around Eq.~\ref{eq:hoop}).} 
	\label{fig:2dmhd-midplanes}
	\end{figure*}

In the HD case, there is a single interface between the disc and stream - at which both RT and KH instabilities would be expected.  With the addition of magnetic field, however, a larger stagnation region develops.  In this case the RT instability would be expected ahead of the stream-edge, at the standoff distance ($y \approx 150$ in Fig.~\ref{fig:2dmhd}) where material feels the largest acceleration.  The KH instability, on the other hand, is still expected at the true stream-edge ($y \approx 250$ in Fig.~\ref{fig:2dmhd}), where the largest velocity shear is present.

\begin{figure}
	\includegraphics[width=\columnwidth]{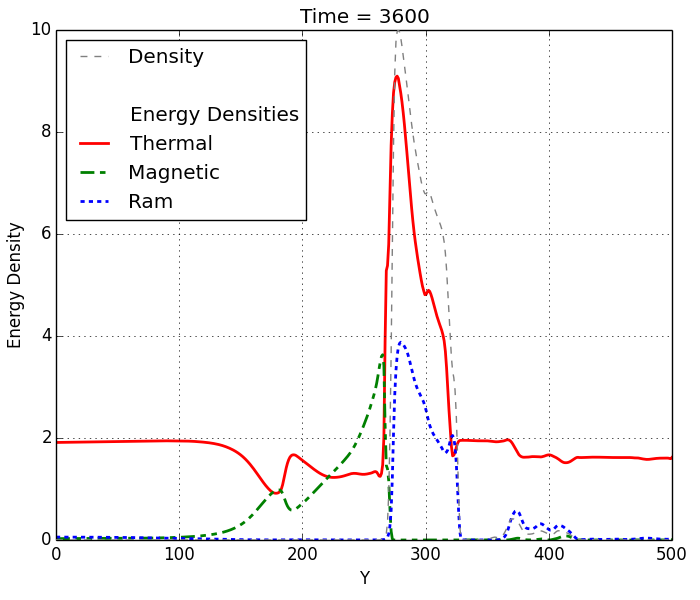}
	\caption{Energy densities in the z-midplane of a 2D MHD simulation.  The stream boundaries are demarcated by the density profile (grey, dashed --- in arbitrary units), and the thermal, magnetic, and ram pressures are over-plotted (red, green, blue).  The roughly $1/r$ profile in magnetic-field strength can be seen leading up to the stream (see the discussion around Eq.~\ref{eq:hoop}).  At this point in the simulation the magnetic field is well-above equipartition just upwind of the stream ($y\approx250$).  The `ram' energy density---calculated as $\frac{1}{2} \rho v_x^2$---seen inside the stream is due to spurious, bulk x-velocity ($v_x$) of the stream (see \tsec{sec:conc}).}
	\label{fig:2dmhd-pres}
	\end{figure}

The nature of the draped (stagnation) region is illustrated in Fig.~\ref{fig:2dmhd-pres} which shows the thermal, magnetic and kinetic (`ram') energy densities in the z-midplane of the 2D MHD simulation.  One repercussion of the softened density contrast between our simulated disc and stream is that the latter acquires a bulk motion from the ram pressure of the disc, accentuated by the increased cross-section which the draped field lines present.  This effect is apparent in Fig.~\ref{fig:2dmhd-pres} which shows the `ram' kinetic energy --- that due to x-velocity ($V_x$), i.e.~$\rho V_x^2$ --- clearly apparent in the stream.  The energy density of the draped magnetic field grows dramatically at the disc--stream interface, rising from an initial $2\%$ of thermal to almost $200\%$---i.e.~twice the equipartition value.
If this were a plane-parallel interface, the magnetic field would push-back against the disc and halt the inflow of additional material as soon as the field strength reaches equipartition.  The curved geometry of the stream, however, allows the magnetic field to be further amplified.
To elucidate the forces at play, it is helpful to write the Lorentz force, $\vec F = \vec j \times \vec B/c$, in a different form. In the disc midplane the Lorentz force in the $\hat y$ direction due to $B_z$ is,
	\begin{equation}
	F_y = -\frac{\partial}{\partial y}\left(\frac{B_z^2}{8\pi}\right) + \frac{B_z^2}{4\pi R_c}, 
	\label{eq:hoop}
	\end{equation}
where $R_c$ is the radius of curvature of the magnetic field lines in the $y-z$ plane \citep{tch08}. The first term on the right-hand side is the pressure gradient force and the second term is the so-called ``hoop stress,'' which comes about due to magnetic field tension. We find that the hoop stress counter-balances most of the outward pressure gradient of the magnetic field, and this causes the magnetic field to increase in strength toward the stream, essentially without any bound, limited only by  the supply of vertical magnetic flux and the stream inertia. This magnetic field is sourced by an electric current, which flows along the surface of the stream. It is thus not surprising that the structure of the magnetic field in front of the stream resembles that around a line current, as can be seen in Figs.~\ref{fig:2dmhd} and \ref{fig:2dmhd-midplanes}: the field strength falls off with increasing distance away from the centre of the stream, roughly following the scaling $B_z \propto R^{-1}$, where $R$ is the distance from the centre of the stream. In fact, in this configuration the pressure gradient term is perfectly balanced out by the hoop stress term. In the simulations, we find a slightly different scaling, $B_z \propto R^{-2/3}$, reflecting the fact that the magnetic field is not axisymmetric around the stream axis.

\begin{figure}
	\begin{center}
	\includegraphics[width=\columnwidth]{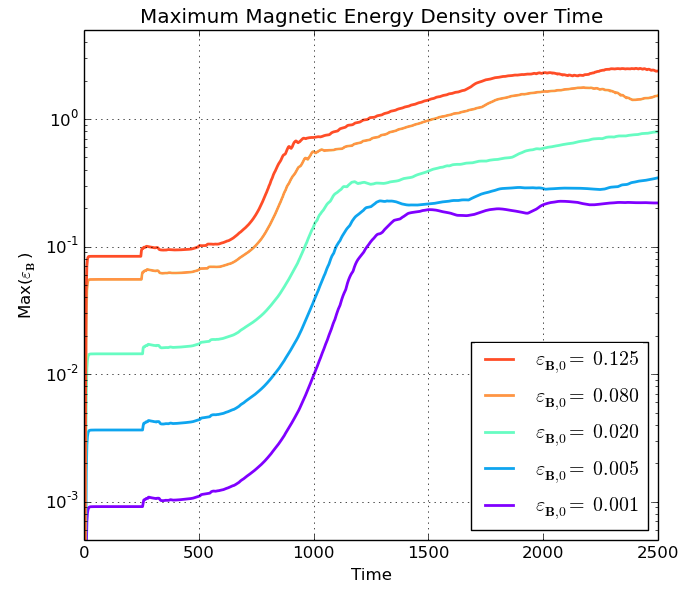}
	\end{center}
	\caption{The maximum magnetic energy density ($\varepsilon_B$) in the grid midplane over time.  Shown are a variety of initial, disc magnetic energy densities ($\varepsilon_{B,0}$).  The first bump at $t\approx 250$ is when the injected disc material (with magnetic field) first reaches the stream, and begins to pileup.  Despite two orders of magnitude variation in the initial field-strength, the piled-up strength only varies by a factor of ten.  This observation could be affected by the finite-mass of the stream (i.e.~the density contrast of only $10^3$).  Note that the field strengths start out just below their `initial values' because of how the magnetized material is being injected into the initially-unmagnetized domain, starting at $t=0$.}
	\label{fig:2dmhd-maxb}
	\end{figure}

Variations in the initial field geometry should have only minor effects on the resulting, concentrated field, because the components of the magnetic field which cross the midplane (i.e.~$B\approx B_z$) will be draped around the stream, while parallel components ($B_x$ and $B_y$) will move past it --- thus selectively amplifying $B_z$.  We consider a quiescent accretion disc threaded by an equipartition strength magnetic field.  To explore the effects of varying field strengths and geometries, we have performed a set of additional 2D MHD simulations.  Fig.~\ref{fig:2dmhd-maxb} shows the temporal evolution of maximum magnetic energy density in the z-midplane, for numerous initial field strengths.  The injected disc material contains magnetic field, while the material initially filling the grid does not.  When the magnetized material first encounters the disc ($t\approx250$ in the simulations), a small bump in field strength is apparent.  After this point, the field rapidly begins to pileup.  Eventually, the stream begins to be driven by the disc along $\hat{y}$, which decreases the rate at which magnetic field accumulates --- causing the turnover between $t\approx900$ (for $\varepsilon_{B,0}=0.125$, red) and $t\approx1400$ ($\varepsilon=0.001$, purple).  Despite two orders of magnitude change in the initial field strength, the accumulated field strength differs by only a factor of ten --- and less-so at the point when the stream begins to move noticeably.  Thus, even with mild variations in the initial field geometry, we can expect the resulting field to be fairly insensitive to those variations \citep{vik01}.

% ====== INSTABILITY =======
\subsection{Disc--Stream Interface Instabilities}
\label{sec:instab}
The stellar debris-stream imbedded in a rotating accretion disc is susceptible to both Rayleigh-Taylor and Kelvin-Helmholtz instabilities.  Rayleigh-Taylor (RT) instabilities develop as a result of a lower-density material, in this case the magnetized disc, being decelerated into a higher density one, here, the stellar stream.  RT sets in for perturbative wavelengths above a critical value $\lambda > \lambda_\trt{c,rt}$ determined by the ratio of acceleration $a$ to surface-tension $T$,
	\begin{equation}
	\label{eq:rt-crit}
	\krt \equiv \frac{1}{\lambda_\trt{c,rt}} = \sqrt{\frac{a(\rho_2 - \rho_1)}{T}},
	\end{equation}
for fluids with lower and higher densities, $\rho_1$ and $\rho_2$, respectively \citep[\textsection{92},][]{cha61}.  

Kelvin-Helmholtz (KH) instabilities, on the other hand, develop for wavenumbers which obey the inequality,
	\begin{equation}
	\label{eq:kh-crit}
	 \kkh T - \frac{a(\rho_2 - \rho_1)}{\kkh} < \frac{\rho_1 \rho_2}{\rho_1+\rho_2}  (v_2 - v_1)^2 ,
	\end{equation}
for a wavenumber $\kkh$, fluid velocities $v_1$ and $v_2$, densities $\rho_1$ and $\rho_2$, and acceleration magnitude $a$ \citep[\textsection{101},][]{cha61}.  In our toy model, the stream and disc velocities are orthogonal, and thus $v_2 - v_1$ is either $v_s$ or $v_d$ depending on the axis of interest.  The strongest acceleration points from the stream into the disc, i.e.~$\vec{a} \approx -|a| \, \hat{y}$ in our simulations.

% ==== Instabilities: Forces ====
\subsubsection{Forces}
\label{sec:forces}
In both cases, surface tension acts to stabilize the interface.  The disc material at the disc--stream interface, contains a strong magnetic field, which contributes an effective surface tension given by, $T_\trt{B} \propto B^2/R_c$, where $R_c$ is again the local radius of curvature.  Only the component of magnetic field in the direction of the perturbation is important as field lines do not resist shearing.  The effective tension provided by the magnetic field is \citep[e.g.~\textsection{106},][]{cha61},
	\begin{equation}
	\label{eq:b-tens}
	T_\trt{Eff} = \frac{B^2}{2\pi} \frac{k_x^2}{k^3} = \frac{B^2}{2\pi} \frac{\cos^2{\theta}}{k},
	\end{equation}
where $\theta$ is the angle between the magnetic field $\vec{B}$ and the perturbation wavevector $\vec{k}$.  Perturbations grow in the plane of the disc--stream interface---i.e. either along the stream's axis ($\hat{x}$), or out of the disc plane ($\hat{z}$).  Components of the field in the $\hat{x}$ direction are easily shed on either-side of the stream, and thus will not tend to buildup.  It can then be expected that the development of instability will be relatively unhindered along the stream axis $\hat{x}$ but suppressed out of the disc-plane $\hat{z}$ (see Fig.~\ref{fig:toy} for the sketch of problem geometry).

While the cold, inviscid stream itself should have negligible surface tension, it may contain some relic, stellar magnetic-field (\tsec{sec:debris}).  Since the stellar material has been stretched out into a stream, the field component out of the plane of the orbit ($\hat{z}$) should be drastically diminished due to flux freezing while the component of magnetic field along the stream-axis may still be important.  This $\hat{x}$ component of the field would tend to suppress the development of instability in the stream material.  Using the stellar magnetic field calculated in Eq.~\ref{eq:b-str}, the effective tension from Eq.~\ref{eq:b-tens} is very roughly,
	\begin{equation}
	\begin{split}
	\label{eq:str-tens}
	T_\trt{Eff} = \, 	&	10^{16} \frac{\tr{erg}}{\tr{cm}^{-2}} \eqsp \bstarz^2 \, \mstarz^{13/4} \, \mbhfive^{-9/4} \, \rrad^{-3} \, \scale{\lambda}{\rstr}.
	\end{split}
	\end{equation}
	
For both RT and KH instabilities, the relevant acceleration is that of hydrodynamic drag, or ram pressure, of the disc interacting with the stream.  For an infinite cylinder, the drag force per unit length $f_l$ is given by,
	\begin{equation}
	f_l = \rho_d v_d^2 R_\trt{Eff} C_\trt{D},
	\end{equation}
for an effective cylinder radius $R_\trt{Eff}$, disc density and velocity $\rho_d$ and $v_d$, and drag coefficient $C_\trt{D}$---which is near unity for a turbulent wake (appropriate for our models).  The effective radius of the cylinder could be much larger than the physical extent of the stream due to the draped-flux region which is what the incoming disc material directly interacts with, as discussed in Sec.~\ref{sec:magnetohydrodynamics} and shown in Fig.~\ref{fig:hd-vs-mhd}.  The ram pressure of the disc is then transmitted to the stream-interface via magnetic pressure and tension.

We parametrize the effective radius as $R_\trt{Eff} = \eta R_{str}$, and consider a fiducial model with the most conservative estimate of $\eta \sim 1.0$.  The acceleration can then be expressed as,
	\begin{equation}
	\begin{split}
	\label{eq:accel}
	a 	&	= \frac{4}{\pi} \frac{\rho_d v_d^2}{\strdens \rstr} \scale{\eta C_D }{1.0}, \\
		&	= 1.3 \tr{ cm s}^{-2} \eqsp \ttime^{5/3} \, \mstarz^{-1/4}\, \mbhfive^{-1/4} \, \rrad^{-2}.
	\end{split}
	\end{equation}

% ==== Instabilities: Critical Scales ====
\subsubsection{Critical Scales}
\label{sec:crit-scales}

Using this acceleration (Eq.~\ref{eq:accel}), and the restoring force from Eq.~\ref{eq:str-tens}, we can calculate the critical RT wavelength using Eq.~\ref{eq:rt-crit},
	\begin{equation}
	\begin{split}
	\label{eq:lrt-crit}
	\lambda_{RT,c} 	&	\approx \frac{\bstr^2}{2\pi a \strdens}, \\
					&	= 1.3 \, \rstr \eqsp \bstarz^2 \,\mstarz^3 \,\mbhfive^{-3/2}\, \rrad^{-3/2}.
	\end{split}
	\end{equation}
RT instability is effective at $\lambda>\lambda_{RT,c}$.
	
Similarly, using Eq.~\ref{eq:kh-crit}, we can find the critical wavelength for KH instability.  Note that after plugging in the expression for $T_\trt{Eff}$ the wavelength dependence of the tension-term drops out.  After making the approximation that $\strdens \gg \rho_d$, the critical wavelength can be expressed as, 
	\begin{equation}
	\begin{split}
	\label{eq:lkh-crit}
	\lambda_\trt{KH,c}	&	\approx \frac{1}{a \strdens} \left[ \frac{\bstr^2}{2\pi} - \rho_d v_s^2 \right].
	\end{split}
	\end{equation}
The first term in Eq.~\ref{eq:lkh-crit} is the same as Eq.~\ref{eq:lrt-crit}, but that constraint is softened significantly by the second term, $\frac{\rho_d v_s^2}{a \strdens}~\approx~4.3~\,~\rstr~\scale{\mbh}{\mfive}$, which is always dominant.  Therefore we expect the Kelvin-Helmholtz instability to be effective at all wavelengths.  These results suggest that unless the magnetic field in the disrupted star was exceptionally large (or in some other way, the stream's field along $\hat{x}$ was significantly enhanced), instability on the scale of the stream-radius and larger should proceed relatively unhindered.

% ==== Instabilities: Growth Rate ====
\subsubsection{Growth Rates}
\label{sec:growrate}
The growth rates of the Rayleigh-Taylor and Kelvin-Helmholtz instabilities, $\omega_\trt{rt}$ and $\omega_\trt{kh}$, are given in Eq.~\ref{eq:rt-growth} and Eq.~\ref{eq:kh-growth} respectively \citep[\textsection{92} \& \textsection{101},][]{cha61}:
	\begin{equation}
	\label{eq:rt-growth}
	\begin{split}
	\omega_{rt} = & \, \left[ \left( \frac{\strdens - \rho_d}{\strdens + \rho_d} \right) a k \right]^{1/2} \approx \sqrt{ a k }, \\
				= &	\, 4.3\E{-5} \textrm{ s}^{-1} \eqsp \mstarz^{-1/4} \, \mbhfive^{-1/2} \, \ttime^{5/6} \, \rrad^{-3/2} \scale[-1/2]{\lambda}{\rstr} ;
	\end{split}
	\end{equation}
	\begin{equation}
	\label{eq:kh-growth}
	\begin{split}
	\omega_{kh} = & \, \frac{\sqrt{\strdens \rho_d}}{\strdens + \rho_d} k v_s \approx k v_s \sqrt{ \frac{\rho_d}{\strdens} }, \\
				= & \, 1.3\E{-4} \tr{ s}^{-1} \eqsp \mstarz^{-1/4} \, \ttime^{5/6} \, \rrad^{-3/2} \scale[-1]{\lambda}{\rstr} .
	\end{split}
	\end{equation}
Both rates give comparable timescales, about $10^4$ s for the inner orbits near the BH, two orders of magnitude shorter than the accretion timescale $\tm \approx 1.1\E{6}$ s.  Near apocentre, the timescale for instability growth becomes larger than $\tm$ when we consider wavelengths ($\lambda$) comparable to the stream radius $(\rstr$) --- which grows linearly with distance from the BH.  Even at apocentre, for instabilities comparable in size to the Schwarzschild radius of our fiducial BH, the instability timescale will still be within a factor of a few of $\tm$.

Both types of instability should be dynamically important, and in particular both phenomena should have time to grow to the scale of the debris stream radius.  Kelvin-Helmholtz perturbations are fundamentally a surface phenomenon, restricted to an outer layer comparable to the wavelength of the perturbation.  Rayleigh-Taylor, on the hand, has no such restrictions, and could lead to disruption of the stream itself.  Note, however, that the above analysis follows the incompressible (subsonic) formalism --- whereas the stream in the tidal disruption is mildly supersonic, as are our simulations (\tsec{sec:ft-simulations}).

There are additional instabilities which could come into play.  In particular, because the magnetic field is wrapped around the stream's axis, it could be susceptible to `kink' and `sausage' instabilities \citep[e.g.][]{lee88}.  Because the draped component of field should always be much stronger than the axial component (the relic, stellar field), the internal field may not be sufficient to stabilize the stream.

This stability analysis of the debris stream suggests that despite its 10 orders of magnitude density excess over the disc, its structure could be significantly disrupted.

	% FIGURE: 2D MHD - Midplane Pressures (5)
	
	% FIGURE: 2D MHD Magnetic Field Growth (6)

% ====== FLUX TRANSPORT =======
\subsection{Flux Transport}
\label{sec:flux-transport}

% CHARACTERISTIC SCALES
\subsubsection{Characteristic Scales}

	% FIGURE: 3D MHD Magnetic Field Lines and Density Slices (7)
	\begin{figure}
	\begin{center}
	\includegraphics[width=\columnwidth]{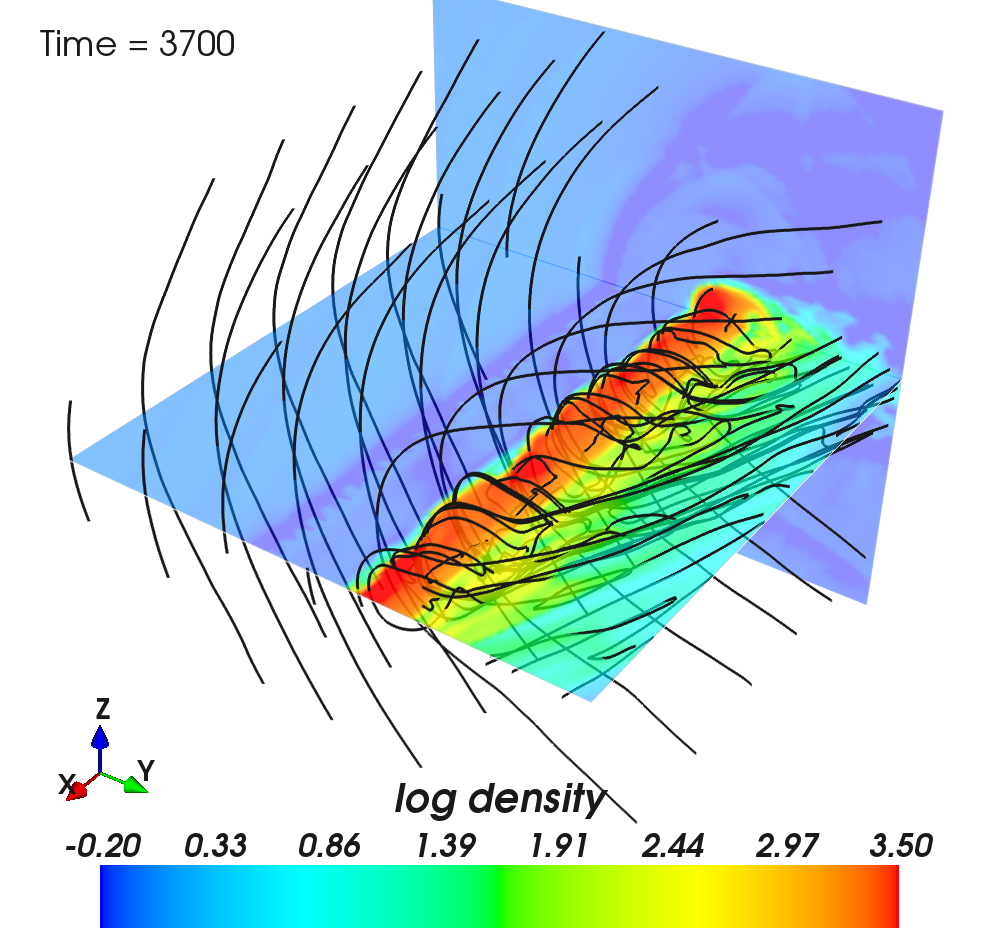}
	\end{center}
	\caption{Slices of log-density in the z-midplane and at the x-edge of a 3D MHD simulation are shown, along with magnetic field streamlines.  At this late time (3700 simulation units), the magnetic field is thoroughly draped over the stellar debris stream -- the boundary of which appears wavy and irregular due to the Kelvin-Helmholtz instability.  `Fingers' of the Rayleigh-Taylor instability are also apparent just upwind (left, $-y$) of the stream.  Magnetic field lines immediately surrounding the stream are dragged along with it ($+x$ direction), and field lines in the down-wind wake of the stream (right, $+y$) can be seen to trail that motion.  \movielink{http://youtu.be/4UbddPLoZeQ} }
	\label{fig:3d}
	\end{figure}

Based on the geometry of the toy-problem setup we can estimate the rate at which magnetic flux can be captured and transported along with a significantly perturbed stream.  We can imagine that some fraction of a magnetic field of strength $B$, within some boundary thickness $\Delta r$, is advected along with the stream's velocity $v_s$.  The rate of flux transport---the amount of magnetic flux transported per unit time---is then,
	\begin{equation}
	\zeta = \xi \, B v_s \Delta r, \label{eq:zetadef}
	\end{equation}
where we have introduced $\xi$ as a coefficient describing the flux capture efficiency.  Equivalently, $\xi$ describes the fraction of flux threading the disc, which after interacting with the debris stream, is carried along with the stream.  This expression for the flux-transport only applies on small-scales (i.e.~those appropriate for the `toy-problem'), as it does not take into account the global structure of the disc, or non-linearity in the flux capture process.  For example, at a particular time and radial distance along the stream, the magnetic field being captured is compounded by the field accumulated from up-stream at previous times.  Nonetheless, this model allows us to numerically calculate the efficiency which can then be applied to the global, physical problem of flux-capture in a tidal-disruption stream.

Our simulations show that the magnetic field strength will not be limited to equipartition values with the ram-pressure of the disc, or its internal pressure.  Instead the magnetic field, supported by hoop stress, will continue to drape around the debris stream growing roughly linearly with time and echoing the inflow of magnetic flux from the disc, i.e.~$B=B_q \cdot \ttime$, where the quiescent magnetic field strength $B_q$ is estimated in Eq.~\ref{eq:mag}.  As the magnetic field piles up in front of the stream, the standoff distance continues to increase over time.  While individual fluid-parcels may not encounter the stream directly for some time, the presence of the draped magnetic flux is communicated to the disc--stream interface by the ever increasing magnetic field strength, and standoff distance.

The thickness of the boundary at a given time could be approximated explicitly using the growth-rate $\omega(\lambda)$, but because the growth timescales are comparable or shorter than the characteristic time of the problem, we will assume that $\Delta r \sim \rstr$.  As soon as a single wavelength exceeds the linear-growth regime it will couple and excite other wavelengths --- thus the fastest (largest) wavelengths are the limiting factor, reinforcing the aforementioned assumption.  The characteristic scale for flux transport is then,
	\begin{equation}
	\begin{split}
	\label{eq:ft-rate}
	\zeta 	\approx	&	\, \xi \,  v_s B_q \ttime \,  \rstr \\
			= 		&	\, 3\E{22} \tr{ G cm}^2 \tr{ s}^{-1} \eqsp \mstarz^{1/4} \, \mbhfive^{1/2} \, \rrad^{-3/4} \, \ttime \, \ximo,
	\end{split}
	\end{equation}
for an efficiency $\xi = \ximo \cdot 10^{-1}$; we remind the reader that here $r = R/\rs$ is dimensionless distance from the BH and $\tau = t/\tm$ is dimensionless time. Naively, this capture rate suggests a flux-accumulation timescale of a few times $10^{6} \textrm{ s}$ to reach the required $\Phi_\tr{jet} \sim 10^{29} \tr{ G cm}^2$ inferred from observations of \sw.  This timescale, comparable to the debris return timescale, is consistent with the onset and evolution of the \sw~event.

While the flux-capture rate in Eq.~\ref{eq:ft-rate} is promising for accumulating the required magnetic field, this is still a \textit{local}, characteristic rate.  A more detailed, \textit{global} calculation, taking the overall quiescent disc structure into account, is presented in Sec.~\ref{sec:global_ft}, and found to agree with the above.

	% FIGURE: 3D MHD Vel All Cuts With Instability (8)
	\begin{figure*}
	\includegraphics[width=1.0\textwidth]{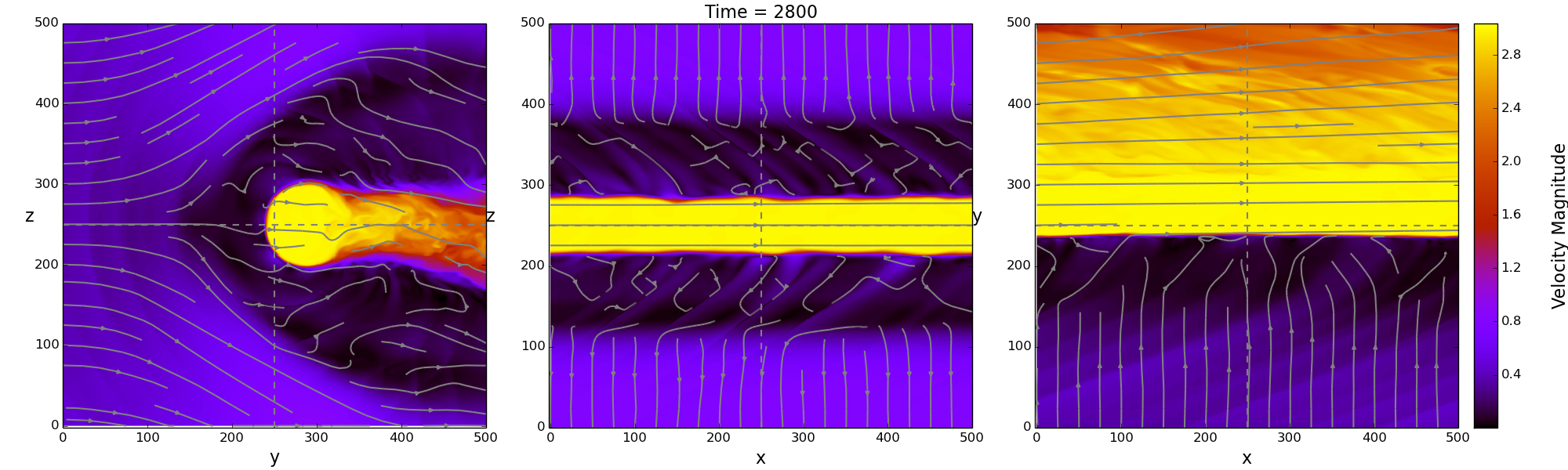}
	\caption{Cross-sections along each midplane of a 3D MHD simulation, showing the structure of $|v|$.  Instabilities are clearly apparent in all directions, with waves propagating out from the disc--stream interface, and conical shocks emanating from the supersonic debris stream.  The velocity-field in the plane of each panel is over-plotted with white streamlines.  Midplane lines are demarcated by grey, dashed-lines -- corresponding to each of the different panels.  \movielink{http://youtu.be/MrCO-KYTRwY} }
	\label{fig:3dmhd-vels}
	\end{figure*}

% SIMULATIONS
\subsubsection{3D MHD Simulations}
\label{sec:ft-simulations}
To demonstrate the efficacy of the schematic flux capture process, and to constrain the efficiencies at which it can occur, we have performed 3D MHD simulations of the toy model described in Sec.~\ref{sec:toy_model}.  As in the 2D case, 
we use the softened density contrast between the stream and the disc to make computations manageable.

A snapshot of our 3D MHD simulation is presented in Fig.~\ref{fig:3d} showing slices of log-density in the z-midplane and at the x-edge of the box, along with magnetic field streamlines.  At this late time (3700 simulation units), the magnetic field is thoroughly draped over the stellar debris stream -- the boundary of which appears wavy and irregular due to the Kelvin-Helmholtz instability.  At this point the magnetic field lines immediately surrounding the stream are comoving with it, and lines in the down-wind wake of the stream (right) can be seen to trail that motion --- more clearly apparent in the animated versions included online (\href{http://youtu.be/lWeJCzDR4Yo}{http://youtu.be/lWeJCzDR4Yo}).

In our fiducial 3D MHD model, we initialize the stream with a velocity $v_s = 3 \, \hat{x}$, at a density $\strdens = 10^3 \rho_d$, where $\rho_d \equiv 1.0$.  Both fluids are initialized with the same pressure, $p_s = p_d = 1.0$.   Periodic boundary conditions are used on the $-\hat{x}$ and $+\hat{x}$ boundaries.  The disc is injected from the $-\hat{y}$ boundary with a velocity $v_d = 0.5 \hat{y}$ and magnetic field $B = B_d \hat{z}$ such that the magnetic energy density $\varepsilon_B \equiv B^2/8\pi = 0.02$.  The three remaining boundaries ($+\hat{y}$, and $\pm\hat{z}$) employ ``diode''-outflow\footnote{A `diode' boundary condition is one which allows an unrestricted outflow, but prevents any inflow of material.  See Appendix~\ref{sec:app_sims}.} boundary conditions.
	
Fig.~\ref{fig:3dmhd-vels} shows midplane cross-sections of the velocity magnitude, $|v|$.  Instabilities are readily apparent in all panels.  The wake region, behind the stream, which we already saw in our 2D simulations, extends throughout the midplane in the downwind region ($y\gtrsim 300$).  While this area is shielded from magnetic field, and thus contributes minimally to flux transport, it could significantly affect the overall accretion onto the BH (see Sec.~\ref{sec:conc}).  
In 3D, 
the flow becomes tangled and turbulent in the draped region, 
as is especially clearly seen in the left panel of Fig.~\ref{fig:3dmhd-vels}.
In front of the stream, at $y\lesssim250$ in Fig.~\ref{fig:3dmhd-vels}, magnetic flux piles up and drapes around the stream. The magnetic field strength grows in time as the disc rotation brings in even more flux towards the stream.
	
	% FIGURE: 3D MHD Flux Transport (9)
	\begin{figure*}
	\includegraphics[width=1.0\textwidth]{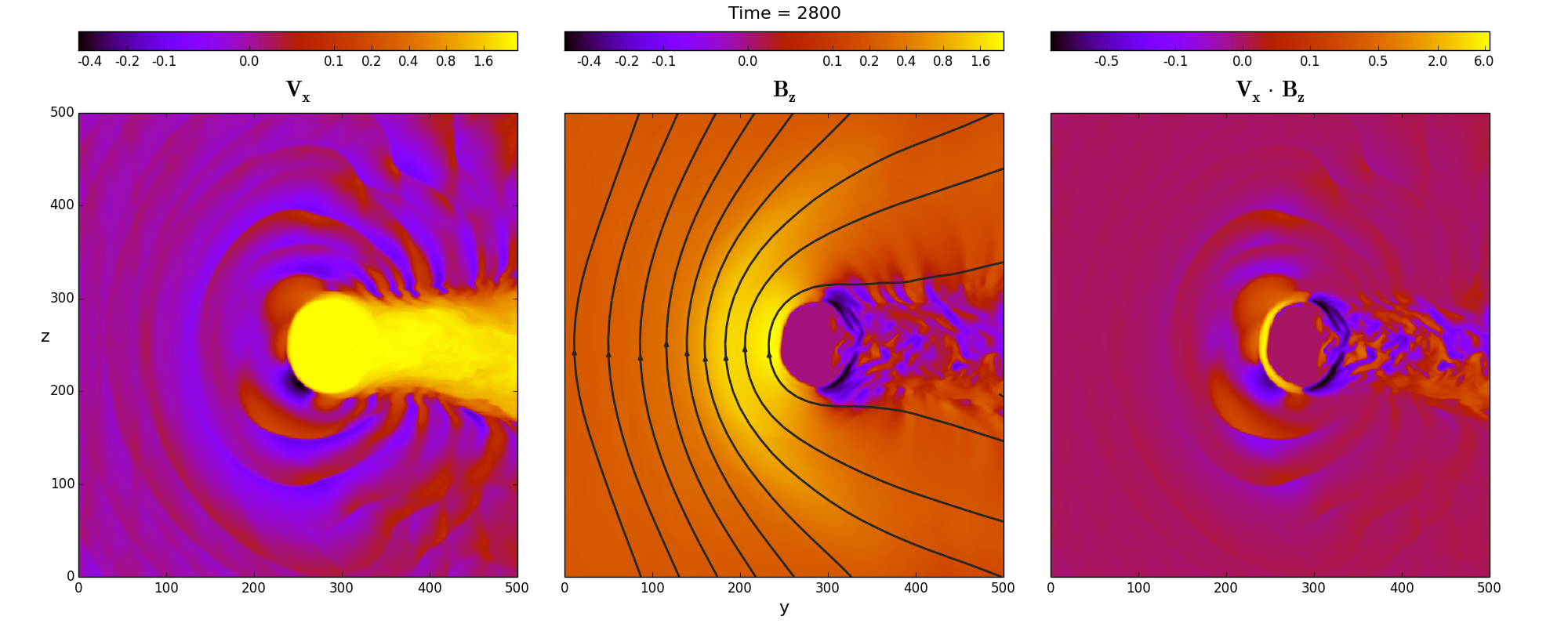}
	\caption{The x-velocity ($V_x$), z-magnetic field ($B_z$), and flux-transport ($V_x \cdot B_z$) per cell at the downstream grid-boundary ($+\hat{x}$) corresponding to the direction of the BH in 3D MHD.  Magnetic field lines are over-plotted (black) in the centre panel.  The velocity field (left panel) shows the presence of conical waves emanating from the super-sonic stream, and the presence of a thin instability region just at the upwind side of the stream.  The magnetic field (centre panel) is strongly amplified as it is draped around the stream.  Due to the instability at the disc--stream interface, the discs magnetic field couples to the stream's velocity and leads to magnetic flux transport by the stream (right panel). This flux transport is seen as the brightened left edge of the stream.  On the right edge of the stream, magnetic field lines are locally inverted --- causing a negative field strength (centre panel) and negative flux-transport (right panel) seen in the darkest purple colors. \movielink{http://youtu.be/Ani8XZ-7lk8} }
	\label{fig:3dmhd-ft-break}
	\end{figure*}
		
\begin{figure*}
	\includegraphics[width=1.0\textwidth]{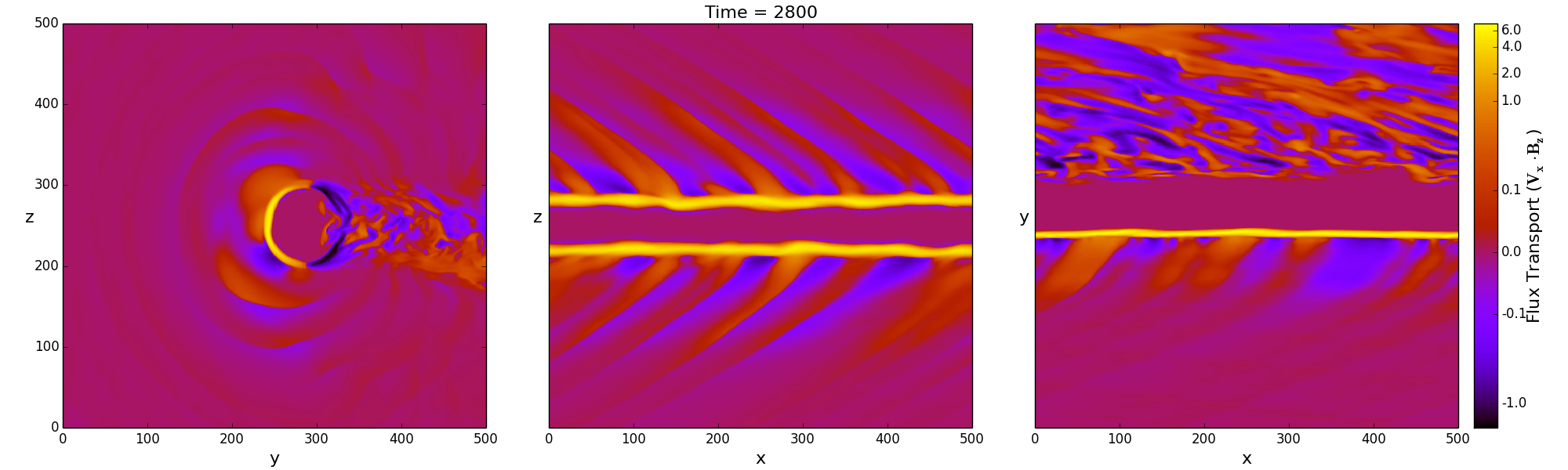}
	\caption{The rate of magnetic-flux transport ($V_x \cdot B_z$) is plotted for each cell in cross-sections of each midplane in our fiducial 3D MHD model.  The dominant contribution to flux-transport is concentrated around a thin boundary region --- where the instability is present --- on the upwind face of the stream.  Conical waves emanating from the fluctuations at the disc--stream interface are also apparent. The instability grows in time, and the size of the flux transport region does so as well, as is apparent in comparison to Fig.~\ref{fig:3dmhd-ft-late}. \movielink{http://youtu.be/-F4aNpoKdUs}}
	\label{fig:3dmhd-ft}
	\end{figure*}

Fig.~\ref{fig:3dmhd-ft-break} shows the velocity in the direction of the stream's motion ($V_x$), the z-component of the magnetic field ($B_z$) and the rate of magnetic-flux transport ($V_x \cdot B_z$), at the grid boundary corresponding to motion towards the BH ($+\hat{x}$).  The highest concentration of draped magnetic field lines are directly upwind (left) of the stream, which shows the largest flux-transport.  A small amount of negative-flux transport is apparent on the top and bottom of the downwind (right) side of the stream, where field lines have been twisted by more than $\pi/2$, and are locally reversed.  The same flux-transport quantity is plotted in Fig.~\ref{fig:3dmhd-ft} for all midplanes, using a broken, symmetric-log\footnote{That is, $-\log{|V_x B_z|}$, below $-0.1$; linear up to $0.1$; and $\log{V_x B_z}$, above.} scaling which highlights the conical waves emanating from the stream.  These waves are launched by oscillations at the disc--stream interface, caused by KH instability.  Flux-transport is again seen to strongly peak along a boundary region surrounding the upwind face of the stream.

	% FIGURE: 3D MHD Flux Transport (10)

\begin{figure}
	\begin{centering}
	\includegraphics[width=0.8\columnwidth]{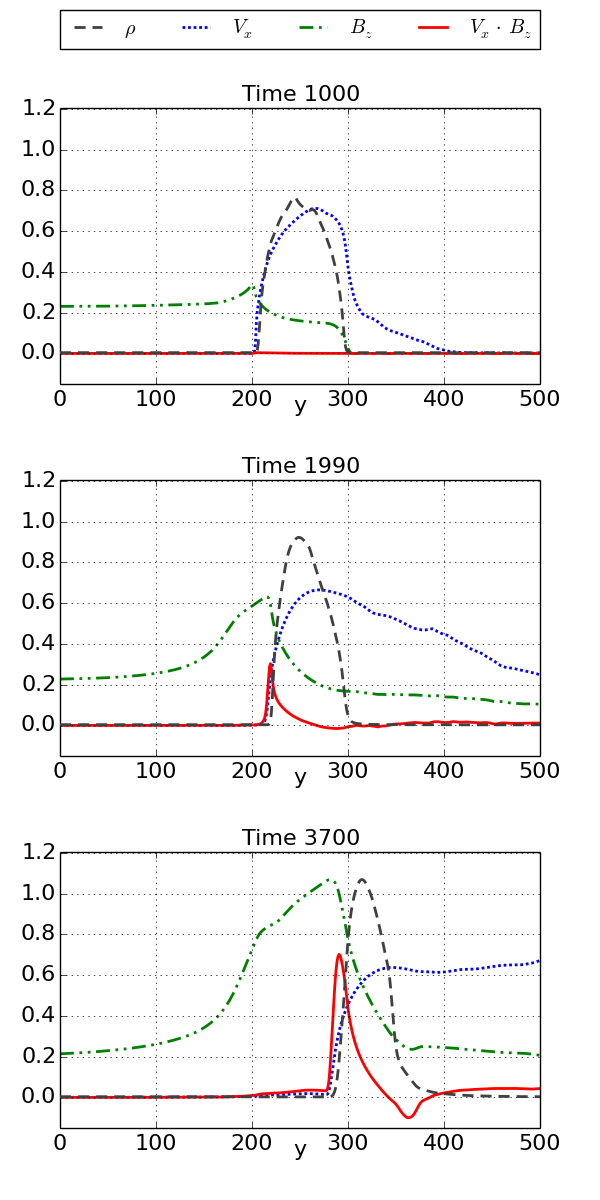}
	\caption{Flux-transport $V_xB_z$ (solid red), $V_x$ (dotted blue) and $B_z$ (dash-dotted green), measured along the $y$-direction, and averaged over both $x$ and $z$.  The projected density is also plotted (dashed black) in arbitrary units.  Three times are shown: the first panel shows the draping of flux over the stream, with no flux-transport.  The rounded structure of the stream (e.g. $V_x$, is due to averaging over the curved structure in the $z$-direction).  The middle panel shows the development of instability, and the beginning of flux-transport at the disc--stream interface.  The boundary region in which transport is effective is fairly small, but still on the order of the stream radius.  Additionally, the magnitude of transport $V_x B_z$ is approximately equal to the average over $V_x$ and $B_z$ independently, suggesting that the velocity and magnetic field are effectively coupled in the boundary layer.  The bottom panel corresponds to a roughly steady-state level of turbulence and flux-transport.  The turbulent wake, formed out of the downwind side of the stream, is clearly apparent - but contributes negligible transport.}
	\label{fig:3dmhd-trans-proj}
	\end{centering}
	\end{figure}

Fig.~\ref{fig:3dmhd-trans-proj} shows one-dimensional projections of the same quantities ($V_x$, $B_z$, and $V_x \cdot B_z$) at three different times.
In these panels, each quantity has been averaged over both $x$ and $z$.  The first panel shows the magnetic field draped around the stream, and increasing in strength all the way to the stream-edge, but the transport $V_x \cdot B_z$ is uniformly zero, as the instability has yet to develop: the parcels of fluid with nonzero $x$-velocity do not coincide with those containing nonzero magnetic field.  The second panel shows the onset of instability at the disc--stream interface, where parcels of magnetized disc material begin to flow with the stream.  Now, the magnetic field couples to the stream's flow, and the flux transport becomes non-zero --- as shown by the solid black-line.  The bottom panel shows s snapshot when the instability has settled into a roughly steady state\footnote{Qualitatively a steady state, but the rate of flux transport (and the associated efficiency $\xi$) increase significantly as the instability grows in size.}.  The value of the flux-transport in Fig.~\ref{fig:3dmhd-trans-proj} shows that the magnetic field couples effectively to the stream, i.e.~the characteristic value of flux-transport ($\approx \langle v_x B_z \rangle$) is comparable to the characteristic values for x-velocity and z-magnetic field independently ($\approx\langle v_x \rangle \langle B_z \rangle$) in the flux-transport region.

	% FIGURE: 3D MHD Flux Transport along Y, projected (11)

	% FIGURE: 3D MHD Flux Transport Efficiency (12)
	\begin{figure}
	\includegraphics[width=1.0\columnwidth,height=0.8\columnwidth]{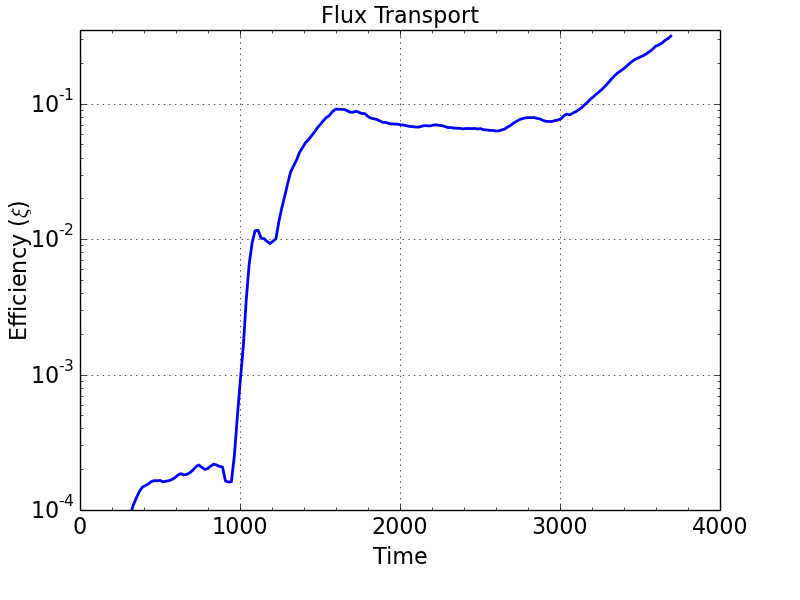}
	\caption{Efficiency $\xi$ of flux-transport calculated in the `toy model' simulations.  Here the rate of flux transport, as a function of time in arbitrary units, is scaled to the measured magnetic field strength ($B$), and stream radius and velocity ($R_\trt{str}$ and $v_\trt{s}$), to yield the effective efficiency computed via Eq.~\ref{eq:ft-rate}.  To correct for the numerical diffusion of magnetic field, the flux-transport is computed as the difference between the flux transport in a 3D simulation and a corresponding 2D simulation.}
	\label{fig:ft-efficiency}
	\end{figure}

% GLOBAL FLUX TRANSPORT
\subsubsection{Global Flux Transport}
\label{sec:global_ft}
In our local simulations, we use a measure of flux transport $\zeta$ (see Eq.~\ref{eq:ft-rate}), to calculate the flux transport in our simulations. For this, we compute flux transport per unit length of the stream by integrating $\zeta$ along the z-midplane of the grid and dividing the result by the $x$-extent of the computational domain.  The details of the capture process are encapsulated in the dimensionless, flux transport efficiency parameter $\xi$, which our simulations give us.  To extract this efficiency measure, the total flux transport is 
first calculated as the sum of cell-by-cell flux transport in the z-midplane at the stream outflow boundary (corresponding to motion towards the BH).  This net value is then normalized to the average magnetic field strength in the flux transport region, the average stream velocity, and the effective stream radius, $\langle R_{\rm str}\rangle$.  $\langle R_{\rm str}\rangle$ is calculated as the standard-deviation of the density-weighted position in the z-midplane; similarly, the `flux-transport region' is defined as within a standard-deviation of the flux-transport-weighted position.

Still, magnetic flux can be transported both by the physical instabilities being studied and due to numerical diffusion of magnetic field into neighboring cells with x-velocity (and vice versa).  We calculate the flux transport in both 3D and 2D simulations.  The 2D simulations are expected to contain the same level of numerical diffusion as in 3D, but none of the physical instability (which is intrinsically a 3D effect).  We correct for the numerical diffusion by subtracting from our 3D results the flux-transport measured in the 2D simulations.  This subtraction is performed on the net flux transport efficiency, described above.  The resulting quantity is a better approximation of the true, physical flux transport.  We have performed convergence studies in both two and three dimensional MHD which have reinforced this approach.  In 3D, increasing the grid resolution increases the flux-transport as instabilities are better resolved.  In 2D, increasing the resolution decreases the flux-transport because the effect of numerical diffusion is being decreased with decreasing cell volumes.  The combination of these trends strongly suggests that we are observing a true, physical flux-transportation process in 3D, and validates the interpretation of 2D results as modeling the numerical diffusion.

Fig.~\ref{fig:ft-efficiency} shows the time-dependence of magnetic field transport expressed in terms of the transport efficiency $\xi$.  At early times, $t\lesssim800$, trace amounts of flux-transport are apparent once the injected magnetic field reaches the stream interface ($t \approx 500$).  The magnetic and velocity fields couple due to weak waves launched from the stream as its boundary relaxes.  As the instability sets in at $t \approx 1000$, transport rapidly increases and the efficiency quickly (by $t=1500$) reaches $\xi \approx 0.1$.  At $t\approx 3000$, the size of the unstable interface region---the flux-transport region---grows to order the stream thickness, at which point the efficiency increases to $\xi \approx 0.5$.  In our simulations, the stream shows spurious bulk motion due to the softened density contrast with the disc.  This motion unfortunately hampers the observation of a saturation, or steady-state at late times. The resulting flux transport efficiency should only continue to increase. 

Fig.~\ref{fig:3dmhd-ft-late} shows a late-time snapshot of density ($\rho$), x-velocity ($V_x$), z-magnetic-field ($B_z$), and flux-transport ($V_x \cdot B_z$) in the z-midplane.  A dashed rectangle is over-plotted to highlight a region where the flux transport is especially apparent.  The density and magnetic field (1st and 3rd from left) panels show that the low-density, highly-magnetized disc material is effectively mixing into the stream.  Once that material mixes in, it is effectively captured by the stream and travels with it, as shown by the x-velocity (2nd from left) panel.  The resulting `flux-transport' ($V_x \cdot B_z$) is shown in the 4th panel.  Here the thickness of the transport region is roughly the stream radius.  Because the strongest magnetic fields are the deepest embedded fields in the unstable region, $\xi$ reaches and remains of order unity.  We expect $\xi$ to remain at this level until either the fossil disc is depleted of magnetic field (i.e.~the disc completes a full rotation---discussed further in \tsec{sec:conc}), or the stream is entirely disrupted.

The measured efficiency represents the fraction of magnetic flux in the vicinity of the stream that is effectively transported along with it.  Using this efficiency value, we can estimate the total magnetic flux that is captured by the debris stream in the physical TDE scenario of \sw\ (see Appendix~\ref{sec:app_global}),
	\begin{equation}
	\label{eq:global-ft-rate}
	\begin{split}
	\Phi_\trt{FC}(t) = 4.4\E{28} \textrm{ G cm}^2 \eqsp \mstarz^{1/4} \, \mbhfive \, \ttime^{1/2} \xihalf,
	\end{split}
	\end{equation}
for an efficiency $\xi = \xihalf \cdot 0.5$.  The corresponding flux capture timescale, to accumulate the required magnetic flux (Eq.~\ref{eq:flux_jet}) is,

	\begin{equation}
	\label{eq:global-ft-time}
	\begin{split}
	\tau_\trt{FC} 	= & \, 3.3\E{6} \textrm{ s} \eqsp \, \lumfe \, \bfac \, \mbhfive^{1/2} \xihalf^{-2}, \\
					= & \, 3.0   \, \tm         \eqsp \, \lumfe \, \bfac \, \mstarz^{-1/2} \xihalf^{-2}.
	\end{split}
	\end{equation}
A few times $10^6$ seconds, or roughly 3 minimum-return timescales of the debris stream, is consistent with the \sw~transient and the fiducial tidal disruption model that we have presented.  In this model, the stream induces the accretion of a significant fraction of the magnetized disc material out to the apocentre of the debris within a few return timescales.  Because the disc is about ten orders of magnitude less dense than the stream, the overall increase in mass-accretion rate due to accreting disc-material is still negligible.

% ==========================================================
% ===================== CONCLUSIONS ========================
\section{Discussion and Conclusions}
\label{sec:conc}
The tidal disruption event \sw~exhibits a stark discrepancy between the properties of the pre-existing system and those inferred from the relativistic jet.  In particular, the magnetic flux threading the disrupted star is entirely insufficient to power the observed outflow. Likewise, the magnetic flux that threaded the BH pre-disruption would be utterly inadequate as well.  Despite this apparent discrepancy, we have used the `afterglow' observations of Swift J1644+57 to show that the pre-existing BH system could contain a sufficient reservoir of magnetic flux stored in the form of a relic accretion disc.  We have demonstrated that the flux from such a fossil disc could be captured, and transported by the infalling, stellar-debris stream.  

In two-dimensional, magnetohydrodynamic simulations, we have found that the magnetic field at the disc--stream interface can be amplified to super-equipartition values due to magnetic hoop-stress.  The drag force, applied from the low-density, magnetized disc material into the high-density stream, leads to Rayleigh-Taylor instability.  At the same time, the large shear between disc and stream velocities destabilizes the interface to Kelvin-Helmholtz instabilities.  

\begin{figure*}
	\includegraphics[width=1.0\textwidth]{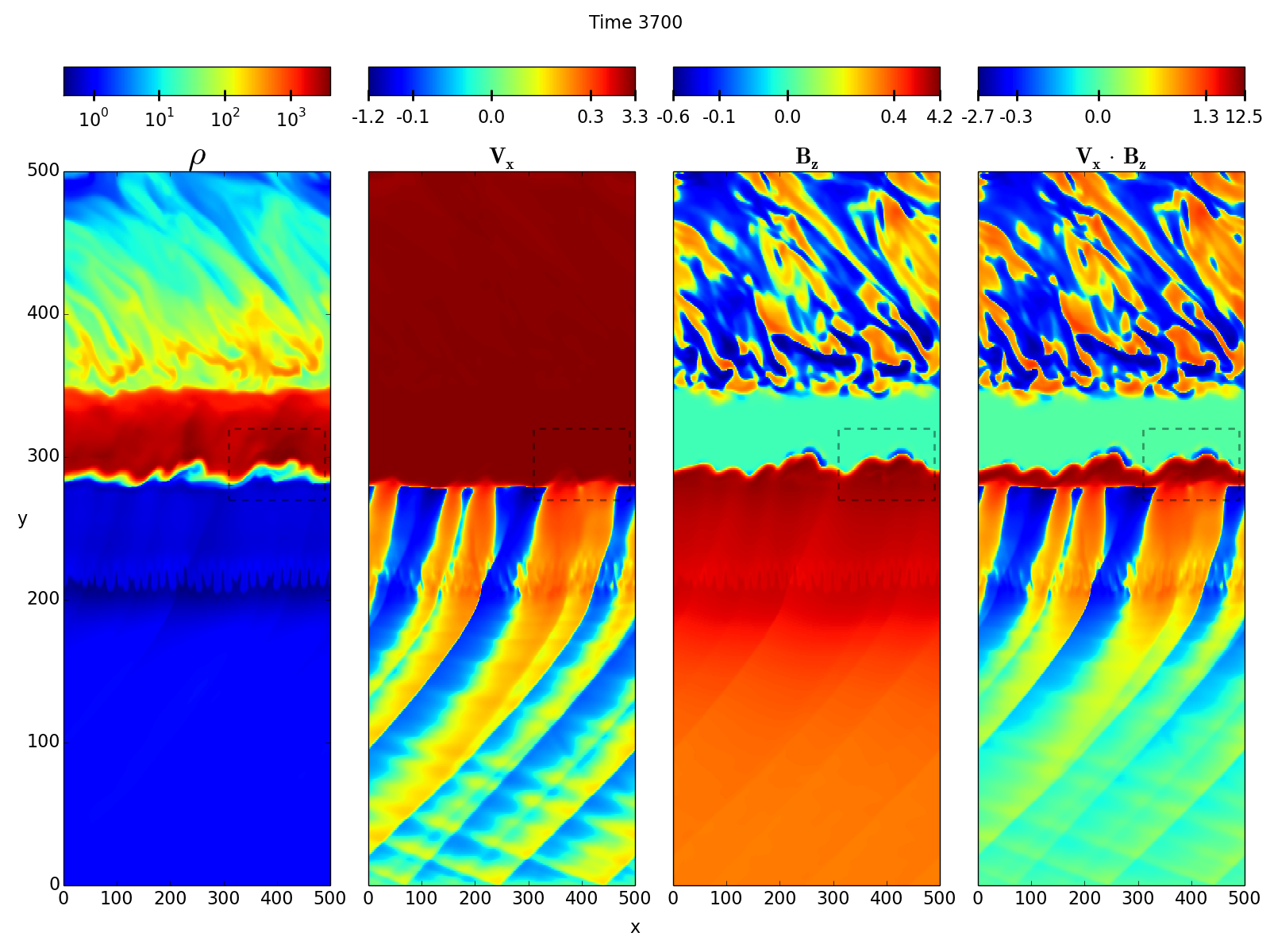}
	\caption{Late-time slices of density, x-velocity ($V_x$), z-magnetic-field ($B_z$), and flux-transport ($V_x \cdot B_z$) in the z-midplane of a 3D MHD simulation.  Both Kelvin-Helmholtz (at $y \approx 290$) and Rayleigh-Taylor ($y \approx 210$) are clearly apparent, especially in density (left-most panel).  The dashed rectangle highlights a region showing that the low-density, magnetized disc material is effectively mixed into the high-density, fast-moving stream --- coupling the magnetic field to the stream's motion.  At late times, the size of this mixing region is approximately the stream radius, leading to a highly `efficient' flux-transport (i.e.~$\xi$ of order unity). Note that the aspect ratio of the panels has been modified from unity for convenience of presentation. Additionally, while the density color-bar is logarithmic, the others are `symmetric-log' with a linear portion around zero (see text for more detail).  \movielink{http://youtu.be/nXPUgQYTjZs} }
	\label{fig:3dmhd-ft-late}
	\end{figure*}

We showed that both types of instabilities have comparable growth rates, which are rapid compared to the dynamical time of the system.  The presence of a large stagnation region ahead of the stream, caused by the draped magnetic field, moves the Rayleigh-Taylor instability to a standoff radius, which is located some distance ahead of the actual disc--stream interface (Fig.~\ref{fig:3dmhd-ft-late}).  The Kelvin-Helmholtz instability is thus the mediator of disc--stream mixing, and the dominant mechanism of magnetic-flux capture.

The rotating disc delivers magnetic field to the interface, where instabilities capture and drag it towards the BH.  We have presented local, three-dimensional simulations of a disc--stream system which tests the efficacy of this scenario.  The simulations demonstrate that instabilities develop at the disc--stream interface.  The presence of those instabilities strongly couples the disc's magnetic field to the stream's motion, leading to flux-capture.  Flux-transport can then be a highly efficient process.  The fiducial model that we have presented suggests that enough magnetic flux can be transported to the central BH via interface instabilities to saturate the hole with magnetic flux. This magnetic flux is dynamically-important: it obstructs the accretion flow and leads to the formation of a MAD following the tidal disruption.  This system is then capable of launching a relativistic outflow of sufficient energetics to power the observed X-ray emission of the \sw~event.

We note that numerous aspects of the preceding analysis have been significantly simplified, and warrant more careful examination.  Simulations that incorporate the global environment are necessary to fully understand  the structure of the tidal-disruption debris streams (see Appendix~\ref{sec:app_stabil}) and its interaction with the existing accretion discs.  In particular, our simulations have not included gravity, nor accounted for radial variations in the disc and debris-stream.  The intricate relationship between the time-varying radial profile of the debris-stream, and how each region of fluid/instability evolves as it flows toward the BH, can only be fully explored through global, large-scale magnetohydrodynamic simulations which include an existing accretion disc, in addition to the returning debris stream.

	% FIGURE: 3D MHD Flux Transport (13)

Our local simulations zoom-in on a small patch of disc--stream interface, where we approximate both the stream and disc as flowing continuously.  In actuality, the inner regions of the disc will complete a full keplerian rotation on much shorter timescales than the stream evolves (i.e.~the minimum return timescale).  This will enforce a maximum amount of magnetic field, which after being draped over the stream, could rebound outwards.  Because the debris stream, with draped field lines, presents a sizable cross-section, a significant fraction of the disc can be stalled.  This material will then lose its centripetal support, and fall back onto the BH.  Disc annuli further out then see a `gap' forming.  A substantial portion of the accretion disc could be depleted in this runaway process.

In addition to global structure, there are additional, more subtle features of TDE dynamics which require further study.  For example, radiative cooling and recombination heating could both be important processes for understanding the detailed structure of the debris stream.  While we have performed additional simulations to confirm that the character of instabilities doesn't change drastically when crossing into trans-, or super-sonic turbulence, simulations with more accurate parameters (especially density contrasts) are needed to obtain more accurate flux-transport measurements.  Our simulations, which have too small of a difference in density between the disc and the stream, artificially enhance the disc's dynamic effects on the stream.  
The analytic calculations of instability growth rates which we have presented (Eqs.~\ref{eq:rt-growth} \& \ref{eq:kh-growth}), use the full, physical density contrasts.  Our flux-transport rates (Eqs.~\ref{eq:ft-rate}--\ref{eq:global-ft-time}), along with our overall conclusions, should not be noticeably affected. 
The bulk motion seen in our simulated debris stream, however, is spurious and hinders the observation of a steady (or saturation) state of flux-transport.

Global simulations will more fully elucidate the saturation state of instabilities, and their eventual effect on the debris stream's structure as it returns to the BH.  Additionally, our results suggest that numerous aspects of the disc--stream interaction could be important for understanding debris-stream circularization and eventual accretion onto the BH -- currently unsolved problems in TDE dynamics.  For example, it is unclear whether purely hydrodynamic effects (e.g.~shocks at pericentre), or the MRI are sufficient to circularize the stream \citep{guil14}.  The wake which we see develop down-wind of the stream, and the capture of the disc's angular momentum, could also be important effects to consider.  If, for example, angular momentum deposited into the debris stream could lead to circularization at larger radii, which could explain the unexpectedly large luminosity of numerous candidate tidal disruption events (e.g.~\citealt{geza12, cho13}---B.~Metzger 2014, private communication).

We are currently exploring these topics by growing our numerical simulations to a complete and self-consistent dynamical model - including a full, rotating accretion disc; central BH; and dynamically evolving debris stream.  Global simulations will be challenging as our convergence studies suggest that while instabilities can be seen as soon as the stream-radius is resolved, smaller length scales---roughly $1/5$ the stream radius---must be resolved to observe physical flux-transport.

Tidal disruption events can be characterized by a simple initial parameter set which describe the star, BH, and the depth of their encounter.  Despite its apparent simplicity, the interaction of debris stream with the pre-existing fossil disc is complex and full of subtlety.  Despite the immense contrast between the density of the tidal debris stream and the fossil disc, the interaction between the two can have pronounced repercussions.  The standard calculation of ballistic orbital trajectories of the debris stream is, perhaps, insufficient to capture the rich dynamics of tidal disruption events.  The topic of tidal disruptions is rapidly growing and can be expected to continue doing so when the next generation all-sky surveys come online \citep[e.g.~LSST;][]{lsst09}, and the rate of TDE detections increases dramatically \citep[see, e.g.~][]{str09}.  The disc--debris stream interaction and the local simulations presented here could be relevant to other types of astrophysical events, for example: gas streams in massive BH binaries discs \citep[e.g.][]{tana13}, or the disruption of hot-Jupiters around stars.

% ====== ACKNOWLEDGMENTS =======
\section*{Acknowledgments}
The authors would like to thank Aleksander S{\c a}dowski, James Guillochon, and Xuening Bai for invaluable discussions on a wide array of topics, including insight into the scientific process.  We are also thankful to Dimitrios Giannios and Brian Metzger for helpful comments and advice in beginning this project.  LZK greatly appreciates the advice and motivation from Edo Berger, Jonathan Grindlay, and Ralph Kraft.
AT was supported by NASA
through Einstein Postdoctoral Fellowship grant number PF3-140115
awarded by the Chandra X-ray Center, which is operated by the
Smithsonian Astrophysical Observatory for NASA under contract
NAS8-03060, and NASA via High-End Computing (HEC) Program through the
NASA Advanced Supercomputing (NAS) Division at Ames Research Center
that provided access to the Pleiades supercomputer, as well as NSF
through an XSEDE computational time allocation TG-AST100040 (AT) and TG-AST080026N (RN) on NICS
Kraken, Nautilus, TACC Stampede, Maverick, and Ranch.  RN was supported in part by NSF grant AST1312651

% =======================================================
% ===================== APPENDIX ========================
\appendix

% ==== Black Hole Mass =====
\section{Mass of the Host Galaxy's Black Hole}
\label{sec:app_bhmass}
The host galaxy for \sw~is unresolved, but the observed B and H-band luminosities suggest a central BH mass of about $2\E{7} \, \msol$ \citep{bur11}.  The authors quote a systematic uncertainty of roughly a factor of 3 --- due to dispersion in the mass relation \citep{marc03}.  Because the galaxy is unresolved, the inferred BH mass tends towards an upper-limit, as more than just the bulge's luminosity is being used.

The X-ray light curve of \sw~may show evidence for variation on timescales around $t_\textrm{var} \sim 78$ s \citep{blo11}, corresponding to the light-crossing time of $M \lesssim 8\E{6} \, \msol$.  At the same time, the X-ray light curve has insufficient signal-to-noise to exclude variability on timescales shorter than tens of seconds \citep{qua12}.

\cite{reis12} reports the possible detection of a quasi-periodic oscillation (QPO) at 4.8 mHz (FWHM around 0.4 mHz), based on detailed followup X-ray observations with Suzaku and XMM newton.  This prospective QPO is not obviously apparent from the Swift data alone.  Assuming this signature is a QPO, associating it with the Keplerian frequency of the ISCO suggests a mass between $5\E{5} \, \msol$ and $5\E{6} \, \msol$ depending on spin.  If the 5 mHz frequency is associated with the light-crossing frequency of a Schwarzschild diameter, it corresponds to a BH mass of $10^7 \, \msol$.  QPOs are usually associated with a disc or inner-corona \citep[e.g.][]{abra01,li03,done07}, while in the \sw~event, the observed X-rays are believed to originate from a jet.  \cite{mcki12} find evidence of a QPO in a simulated jet magnetic field itself with a (spin dependent) frequency $\nu_\trt{QPO,J} \approx c/(70 r_g)$.  Scaling this relation to match the \sw~candidate-QPO suggests a BH mass of about $6\E{5} \, \msol$.

In this study we use a fiducial BH mass of $\mbh = \mbhfive \cdot \mfive$, which we find to be the most consistent with observations, as was also established in \tchek.  While being somewhat on the lower-mass end, we believe this value to be entirely consistent with observations and inferences (e.g.~\tsec{sec:sw}).

% ==== Stream Stability ====
\section{Debris Stream Stability and Structure}
\label{sec:app_stabil}
Typically, each parcel of gas composing the debris stream is assumed to be on independent, ballistic trajectories until it returns to pericentre.  At that point, the converging stellar material interacts hydrodynamically and eventually circularizes \citep[e.g.][]{koch94}.  Even if hydrodynamic effects are negligible far from pericentre, self-gravity may not be.  Based on the density of the stream (Eq.~\ref{eq:tde-dens}), the self-gravity dynamical-time is a few times $10^4$s --- far shorter than the orbital time.

To resist collapse, the stream (still modeled as a uniform cylinder) requires a pressure support of roughly,
	\begin{equation}
	\begin{split}
	p_{sup} =	& \, \pi G \rho^2 R^2 \\
			= 	& \, 1.7\E{7} \frac{\tr{erg}}{\tr{cm}^{3}} \,\, \mstarz^{1/2} \, \mbhfive^{-5/2} \, \rrad^{-1} \, \ttime^{-10/3} .
	\end{split}
	\end{equation}
Near pericentre, the stream is stable to self-gravity, but as distance increases the stream's pressure (Eq.~\ref{eq:str-pressure}) drops much more rapidly than density, creating a critical distance at which the stream becomes vulnerable to collapse.  This critical distance is,
	\begin{equation}
	r_{crit} \approx 245 \, \rs \eqsp \ttime^{10/27} \, \mstarz^{-1/3} \, \mbhfive^{-5/9},
	\end{equation}
which is significantly smaller than the characteristic orbital scales.  The temperature of the stream is roughly,
	\begin{equation}
	\label{eq:str-temp}
	T_{str} \approx 6.1\E{4} \textrm{ K} \eqsp \mbhfive^{-4/3} \, \rrad^{-1} \, \ttime^{-10/9},
	\end{equation}
which is very cold compared to the circumnuclear material (see \tsec{sec:qui-bh}).  Heating from recombination may be important \cite[e.g.][]{koch94}, and if so, would help stabilize the stream against gravity.

Additionally, the shear between fluid elements on slightly different trajectories could also bolster the debris stream against collapse.  Near apocentre, the stream should have a sound speed $c_\tr{s,str} \approx 10^4$--$10^5 \textrm{ cm s}^{-1}$.  Modeling the stream as rotating with an angular velocity $\Omega = 2\pi / \tm$, and having a surface density $\Sigma = \mstar / (R_{str} \cdot \mathcal{L}_\trt{orb})$---where the stream is spread over an area\footnote{$R_{str} \cdot \mathcal{L}_\trt{orb}$ ends up being comparable to the full ellipse area, $\approx \major \minor$.} determined by the stream radius $R_{str}$ and orbital circumference $\mathcal{L}_\trt{orb}$---we can calculate the Toomre parameter \citep{toom64,binn08}, $Q \equiv \, c_\tr{s,str} \, \Omega / (\pi \, G \, \Sigma)$.  At apocentre and after a time $\tm$, to an order of magnitude,
	\begin{equation}
	Q \approx \, 0.3,
	\end{equation}
suggesting that rotation could \textit{marginally} stabilize the debris stream -- especially in lower surface density (more peripheral) areas.  Additionally, while the magnetic field in the stellar debris should be small compared to that of the relic accretion disc (see \tsec{sec:debris}), the magnetic pressure would also act in opposition to gravity.  

We expect self gravity to effect a mild compression of the stream, subdominant to the ballistic dynamics of its evolution.  Fully understanding the internal dynamics of the debris stream would require large-scale, global simulations of its evolution.  Throughout this work, we estimate the stream density using a constant solid angle, i.e.~assuming that self-gravity is sufficiently balanced by pressure, shear, and magnetic fields \citep[see however;][]{guil14}.

% ====== SIMULATION STABILITY =======
\section{Modifications to the Numerical Scheme for Handling High Density Contrasts}
\label{sec:app_sims}
While attempting to simulate the interaction of the disc and the stream with a high density contrast, we ran into numerical difficulties and made modification to the numerical scheme to avoid them.  We found the following three aspects of our toy model to be especially challenging to accurately model using \mbox{(magneto-)hydrodynamic} codes.  First, the stark density contrast between the stream and disc is very difficult to resolve in 3D, even using static or adaptive mesh refinement.  The interface between disc and stream has proven to be susceptible to large flux errors which lead to huge energy-depositions in single cells (`explosions') at or near large density contrasts.  We have observed this phenomenon using numerous combinations of integrators, Riemann-solvers, and hydrodynamic packages \citep[e.g.][]{mign07,sto08}.  Additionally, the supersonic motion of the stream tends to launch shocks and is susceptible to heavy dissipation.  Finally, the super-equipartition strength magnetic fields (i.e.~$B^2 \gg P$), and very cold (i.e.~$P \ll \rho v^2$) stream tend towards numerical errors driving the internal energy (temperature, pressure, etc.) to negative values.

All of the numerical simulations we present here were performed with the Athena MHD code \citep[v4.2;][]{sto08}, using the Corner Transport Upwind (CTU) integrator \citep{gard08}, and the `HLLD' and `Roe' Riemann solvers for MHD and HD respectively \citep[see,][]{sto08}.  While we found that artificially decreasing the integration time-step (using a small Courant factor, $\mathcal{C} \approx 10^{-2}$ to $10^{-4}$) was the most effective way at preventing spurious energy depositions or negative internal energies, the required computational time to complete such simulations was intractable.  Instead, we found the following combination of multiple strategies to be effective.  First, softening the parameters used for our simulations --- in particular, decreasing the density contrast between the disc and stream.  Second, initializing the simulation with a static and zero-magnetic-field background (disc material), and only injecting moving, magnetized material afterwards.  Third, and similarly, starting simulations with a smaller Courant factor ($\approx 10^{-2}$) and increasing it after some time (e.g.~roughly a sound-crossing time) to a more reasonable value (generally $\mathcal{C} \approx 0.1$ to $0.3$).  Fourth, we found that implementing consistent density and pressure floors, i.e.~artificially enforcing both density and pressure to remain above predefined (low) values, significantly improved the code performance.

To implement these floors, we added a function to Athena which, after every integration step, would reset densities and pressures to be above the floors if they were found to be below them.  Typically the floors would be set to roughly an order of magnitude less than the minimum expected values.  For example, for a background (disc) density, $\rho = 1.0$, a floor value of $\rho_\trt{flr} = 0.1$ was effective, and consistently below density values which would typically arise in the grid.  These floors would only be applied very sporadically --- generally in only a few cells, for a few time steps.  Thus the errors which were causing cells to reach non-physical values seemed to be highly transient and localized.  At no time did floors contribute to a substantial, global deviation from mass or energy conservation.  An alternative to implementing pressure-floors is to track additional fluid parameters (e.g.~internal energy or entropy, instead of total energy) which can be `fallen-back upon' if the internal energy (for example) is found to be non-physical.  This will again lead to a local deviation from total-energy conservation, but will do so in a less \textit{ad-hoc} manner.  For simplicity, and for consistency with the density, we opted for flooring.

Despite our modifications, a small `mini-explosion' still occurs early-on in one of our 3D MHD simulations which was used in our final analysis.  This explosion can be seen at early times ($\approx 80$ simulation units) in the animations included with the online version (and on \href{http://youtu.be/4UbddPLoZeQ}{youtube}) on the downwind ($+y$) side of the stream.  We are confident that the effects of this disturbance are negligible (i.e.~when compared with the initial pressure perturbations we insert manually), and only mention it for completeness.

As mentioned in \tsec{sec:ft-simulations}, we have also implemented so-called, `diode' boundary conditions (BCs) in place of the built-in, Athena `outflow' BCs.  An `outflow' condition simply copies the content (density, velocity, etc.) of grid-cells at the domain boundary into the `ghost' or `guard' cells farther out - which are used to calculate gradients at the boundaries.  If a terminal cell develops an inward velocity profile (which is allowed), that velocity will be duplicated into the corresponding ghost-cell.  In this way, information is allowed to propagate into the grid-domain, without being well-defined at the boundary.  The grid can easily be contaminated, or even dominated, by spurious values when this happens.  A more robust procedure is to allow unrestricted outflow when boundary cells have an outward velocity, but fix the ghost-cell velocity to zero --- to restrict inflow --- if the velocity tends inward.  Such a `diode' BC is especially appropriate when material will generally flow out through a boundary, but transient eddies, for example, may occasionally tend to cause an inflow.

	% FIGURE: 3D MHD Flux Transport Over Time (C1)
	\begin{figure*}
	\includegraphics[width=0.8\textwidth]{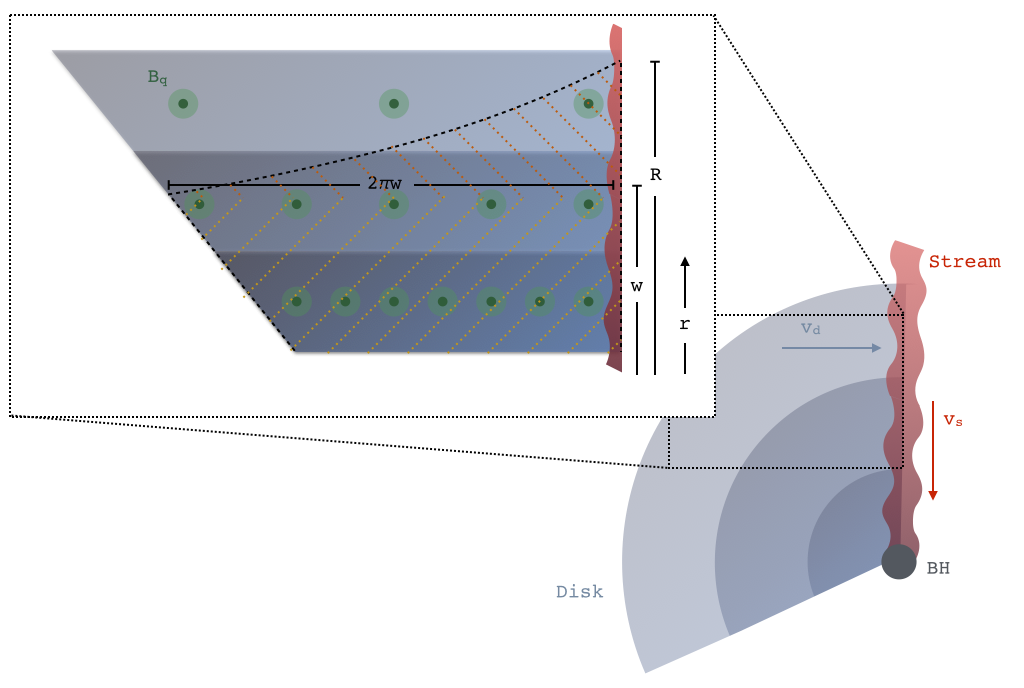}
	\caption{The section of disc which contacts the stream within some time $T$ is shown as the hashed region of the zoom-in.  A parcel of fluid which starts at $R$ is able to reach the BH after $T$.  Such a parcel encounters all of the material in the hashed-region as it rotates into the stream.  The lower hashed-region is for parts of the disc which have completed a full rotation within time $T$ --- i.e. all disc material, out to a radius $w$, has come in contact with the stream.  The upper hashed-region denotes the portion of the outer disc which also contacts the stream.  The entire hashed-region represents the magnetic flux available for capture after $T$.  Simulations yield the capture efficiency, $\xi$, which corresponds to the fraction of the available material which can effectively be transported to the BH.}
	\label{fig:global-ft}
	\end{figure*}

% ====== GLOBAL FLUX CAPTURE =======
\section{Global Flux Capture Calculation}
\label{sec:app_global}
Our `toy model', shown schematically in Fig.~\ref{fig:toy}, represents a small, localized region of the global tidal-disruption event.  The following section describes one way of applying the efficiency calculated in the toy model simulations to estimate the total amount of magnetic flux which can be transported globally.  The total section of disc which comes in contact with the stream at some time $T$, after the tidal disruption, is shown schematically in Fig.~\ref{fig:global-ft}.  After this period of time, the furthest fluid element which is accreted down to the BH, along with the stream, started at some initial radius $R$ from the BH.  In other words, a fluid-parcel which started at $R$, reaches the BH after a time $T$.  The initial radius $R$ can be related to the contact time by solving the inflow equation for the stream, $dr = v_s(r) \, dt = c \left(r/r_s\right)^{-1/2} dt$, i.e.

	\begin{equation}
	\begin{split}
	r(t) = r_s^{1/2} \left[ \frac{3}{2} c t \right]^{2/3},
	\end{split}
	\end{equation}
with $R\equiv r(t=T)$.

The contacted area can be broken down into lower and upper sub-regions.  The lower region is defined as radii for which the entire annulus comes into contact with the stream; i.e.~when the orbital period, $2\pi r / v_d(r)$ is less than the contact time $T$.  This lower region extends from the BH to some radius $l$.  We can parametrize the ADAF rotational velocity profile as $v_d(r) = s \cdot \left( r/r_s \right)^{-1/2}$, and then solve for the limiting radius as,

	\begin{equation}
	\begin{split}
	l = r_s^{1/3} \left[ T \, s \right]^{2/3}.
	\end{split}
	\end{equation}

The upper region, $l < r < R$, has a width-function $w(r)$ determined by the time it takes material to inflow from the initial radius $R$ to the corresponding point $r$, i.e.

	\begin{equation}
	\begin{split}
	w(r)	= & v_d(r) \int_R^r \left[ v_s(r') \right]^{-1} dr', \\
	  		= & \frac{2s}{3c} \frac{ R^{3/2} - r^{3/2}}{r^{1/2}}.
	\end{split}
	\end{equation}

If we parametrize the quiescent ADAF magnetic field as $B_q(r) = B_0 \left( r/r_s \right)^{-5/4}$, then the total magnetic flux in the lower region ($\Phi_l$) can be calculated as,

	\begin{equation}
	\begin{split}
	\Phi_l(T)	= & 2\pi B_0 \int_0^{l(T)} r \left( \frac{r}{r_s} \right)^{-5/4} dr, \\
	  			= & \frac{8\pi}{3} B_0 \, r_s^{5/4} \, l^{3/4}.
	\end{split}
	\end{equation}

The magnetic flux in the upper region can be calculated as,

	\begin{equation}
	\begin{split}
	\Phi_u(T)	= & \frac{2s}{3c} B_0 \, r_s^{5/4} \int_l^R \frac{ R^{3/2} - r^{3/2}}{r^{7/4}}, \\
	  			= & \frac{8s}{9c} B_0 \, r_s^{5/4} \frac{ \left(R^{3/4} - l^{3/4} \right)^2 }{ l^{3/4} }.
	\end{split}
	\end{equation}

Based on the ADAF model, we can solve for the ratio,

	\begin{equation}
	\frac{R}{l}	= \left[ \frac{3c}{2s} \right]^{2/3} \approx 2.3,
	\end{equation}
and see that $\Phi_u(T) \propto B_0 \, r_s^{5/4} \, l^{3/4}$ has approximately the same scaling as $\Phi_l(T)$, but is almost an order of magnitude less.  We can then take the global flux which comes in contact with the stream after a time $T$ as roughly,

	\begin{equation}
	\begin{split}
	\Phi_\trt{cont}(T)	= & \frac{8\pi}{3} B_0 \, r_s^{5/4} \, l^{3/4}, \\
						= & \frac{8\pi}{3} B_0 \, r_s^{3/2} \left[ T \, s \right]^{1/2}.
	\end{split}
	\end{equation}

The global flux transport rate is simply $\zeta_g = \xi \, d\Phi_\trt{cont}/dt$, and the total flux captured is,

	\begin{equation}
	\begin{split}
	\Phi_\trt{FC}(t) = \frac{8\pi}{3} B_0 \, \xi \, r_s^{3/2} \left[ t \, s \right]^{1/2}.
	\end{split}
	\end{equation}

In physical units appropriate for the TDE problem, this can be expressed as,

	\begin{equation}
	\begin{split}
	\Phi_\trt{FC}(t) = 4.4\E{28} \textrm{ G cm}^2 \eqsp \mstarz^{1/4} \, \mbhfive \, \ttime^{1/2} \xihalf.
	\end{split}
	\end{equation}
where the efficiency is taken as $\xi = \xihalf \cdot 0.5$, based on the results from our simulations (\tsec{sec:global_ft}).

\newpage
\bibliographystyle{mn2e}
\bibliography{ResearchExamPaper}

\label{lastpage}

\end{document}